\documentclass[english,pre,twocolumn,superscriptaddress,preprintnumbers,amsmath,amssymb,showpacs,showkeys]{revtex4}
\usepackage[T1]{fontenc}
\usepackage[latin9]{inputenc}
\usepackage{float}
\usepackage{graphicx,color}
\usepackage{subfigure}
\usepackage{epstopdf} 
\makeatletter
\@ifundefined{textcolor}{}
{%
 \definecolor{BLACK}{gray}{0}
 \definecolor{WHITE}{gray}{1}
 \definecolor{RED}{rgb}{1,0,0}
 \definecolor{GREEN}{rgb}{0,1,0}
 \definecolor{darkgreen}{rgb}{0,0.6,0}
 \definecolor{BLUE}{rgb}{0,0,1}
 \definecolor{CYAN}{cmyk}{1,0,0,0}
 \definecolor{MAGENTA}{cmyk}{0,1,0,0}
 \definecolor{YELLOW}{cmyk}{0,0,1,0}
 }

\newcommand{\figurewidth}{\columnwidth}
\newcommand{\beq}{\begin{eqnarray}}
\newcommand{\eeq}{\end{eqnarray}}
\newcommand{\Tr}{{\text Tr}}
\newcommand{\e}{{\text e}}
\newcommand{\rmd}{{\text d}}

\usepackage{amsthm}

\makeatother

\usepackage{babel}

\makeatother

\usepackage{babel}
\usepackage{graphicx}
\usepackage{amsfonts}
\usepackage{amsmath}

\makeatother

\usepackage{babel}

\begin{document}

\title{The performance of the quantum adiabatic algorithm \\ on random instances of two optimization problems on regular hypergraphs}

\author{Edward Farhi}
\affiliation{Center for Theoretical Physics, Massachusetts Institute of Technology, 77 Massachusetts Avenue, 6-304, Cambridge, Massachusetts 02139, USA}
\author{David Gosset}
\affiliation{Department of Combinatorics \& Optimization and Institute for Quantum Computing, University of Waterloo, 200 University Avenue West, Waterloo, Ontario, Canada N2L 3G1}
\author{Itay Hen}
\affiliation{Department of Physics, University of California, Santa Cruz, California 95064, USA}
\author{A. W. Sandvik}
\affiliation{Department of Physics, Boston University, Boston, Massachusetts 02215, USA}
\author{Peter Shor}
\affiliation{Department of Mathematics, Center for Theoretical Physics and CSAIL, Massachusetts Institute of Technology, Cambridge, Massachusetts 02139, USA}
\author{A.~P.~Young}
\affiliation{Department of Physics, University of California, Santa Cruz, California 95064, USA}
\author{Francesco Zamponi}
\affiliation{Laboratoire de Physique Th\'eorique, UMR 8549, CNRS and Ecole Normale Sup\'erieure, 24 Rue Lhomond, FR-75231 Paris Cedex 05, France}
\date{\today}

\begin{abstract}
In this paper we study the performance of the quantum adiabatic algorithm
on random instances of two combinatorial optimization problems, 3-regular
3-XORSAT and 3-regular Max-Cut. The cost functions associated with these two
clause-based optimization problems are similar as they are both defined on
3-regular hypergraphs.  For 3-regular 3-XORSAT the clauses contain three variables
and for 3-regular Max-Cut the clauses contain two variables. The quantum
adiabatic algorithms we study for these two problems use interpolating
Hamiltonians which are stoquastic and therefore amenable to sign-problem free quantum Monte
Carlo and quantum cavity methods. Using these techniques we find
that the quantum adiabatic algorithm fails to solve either of these problems
efficiently, although for different reasons.
\end{abstract}

\maketitle

\section{Introduction}
\label{sec:intro}
The Quantum Adiabatic Algorithm (QAA) ~\cite{farhi_long:01}  is an algorithm for solving optimization problems using a quantum computer. The optimization problem to be solved is defined by
a cost function which acts on $N$ bit strings. The computational task is to
find the global minimum of the cost function. 

To use the QAA, the cost function is first
encoded in a quantum Hamiltonian $H_P$ (called the `problem Hamiltonian') that
acts on the Hilbert space of $N$ spin $\frac{1}{2}$ particles. The problem
Hamiltonian is written as a function of $\sigma_z$ Pauli-matrices and is
therefore diagonal in the computational basis. The ground state of $H_P$
corresponds to the solution (i.e., lowest cost bit string) of the optimization
problem. 

To find the ground state of the problem Hamiltonian, the system is first prepared in the ground state
of another Hamiltonian $H_B$, known as the beginning Hamiltonian.  The
beginning Hamiltonian does not commute with the problem Hamiltonian and must
be chosen so that its ground state is easy to prepare.  Here we use the
standard choice
\[
H_B=\sum_{i=1}^{N}\frac{\left(1-\sigma_x^{i}\right)}{2},
\]
which has a product state as its ground state.

The Hamiltonian of the system is slowly modified from $H_B$ to
$H_P$. Here we consider a linear interpolation between the two Hamiltonians
\begin{equation}
\hat{H}(s)= (1-s)H_B+s H_P \,,
\label{eq:interp}
\end{equation}
where $s(t)$ is a parameter varying smoothly with time,
from $s(0)=0$ to $s(\mathcal{T})=1$ at the end of the algorithm after a total evolution time $\mathcal{T}$.

If the parameter $s(t)$ is changed slowly enough, the adiabatic theorem of
Quantum Mechanics~\cite{kato:51,messiah:62,amin:09b,jansen:07}),
ensures that the system will stay close to the ground state
of the instantaneous Hamiltonian throughout the evolution. After time
$\mathcal{T}$ the state obtained will be close to the ground state of $H_P$.
A final measurement of the state in the Pauli-$z$ basis then produces the
solution of the optimization problem.

The runtime $\mathcal{T}$ must be chosen to be large enough so that the adiabatic
approximation holds: this condition determines the
efficiency, or complexity, of the QAA.  A condition on $\mathcal{T}$ can
be given in terms of the eigenstates $\{ | m
\rangle \}$ and eigenvalues $\{E_m \}$ of the Hamiltonian $H(s)$,
as~\cite{wannier:65,farhi:02}
\begin{equation}
\mathcal{T} \gg \hbar \, { \textrm{max}_{s} |V_{10}(s)| \over
(\Delta E_{\textrm{min}})^2} \,,
\end{equation} 
where $\Delta E_{\textrm{min}}$ is the minimum of the first
excitation gap
$\Delta E_{\textrm{min}} = \textrm{min}_{s} \Delta E$
with $\Delta E = E_1-E_0$,
and $V_{m 0} = \langle 0 | {\textrm d} H / {\textrm d} s | m\rangle$.

Typically, matrix elements of $H(s)$ scale as a low polynomial of the
system size $N$, and the question of whether the runtime is
polynomial or exponential as a function of $N$ 
therefore depends on how the minimum gap $\Delta
E_{\textrm{min}} $ scales with $N$. If the gap becomes
exponentially small at any point in the evolution, then the computation
requires an exponential amount of time and the QAA is inefficient.  The
dependence of the minimum gap on the system size for a given problem is
therefore a central issue in determining the complexity of the QAA. 

A notable feature of the interpolating Hamiltonian \eqref{eq:interp} is that
it is real and all of its off diagonal matrix elements are non-positive.
Hamiltonians which have this property are called stoquastic \cite{bravyi:08}. There is complexity-theoretic evidence
that some computational problems regarding the ground states of stoquastic
Hamiltonians are easier than the corresponding problems for more general
Hamiltonians \cite{Bravyi:09}. It may be the case that quantum adiabatic
algorithms using stoquastic interpolating Hamiltonians (such as the ones we
consider here) are no more powerful than classical algorithms--this remains an
intriguing open question.

An interesting question about the QAA is how it performs on
``hard'' sets of problems -- those for which all known algorithms take an
exponential amount of time.  While early studies of
the QAA done on small systems ($N \leq 24$)~\cite{farhi_long:01,hogg:03}
indicated that the time required to solve one such problem might scale
polynomially with $N$, several later studies using larger system sizes gave
evidence that this may not be the case.

References ~\cite{farhi:02,farhi:08} show that adiabatic algorithms will fail
if the initial Hamiltonian is chosen poorly. Recent work has elucidated a more
subtle way in which the adiabatic algorithm can
fail~\cite{altshuler:09b,amin:09,farhi-2009,altshuler:09,FSZ10}. 
The idea of these works is that a very
small gap can appear in the spectrum of the interpolating Hamiltonian due to
an avoided crossing between the ground state and another level corresponding
to a local minimum of the optimization problem. The location of these avoided
crossings moves towards $s=1$ as the system size grows. They have been called
``perturbative crosses'' because it is possible to locate them using low order
perturbation theory. Altshuler {\it et al.}~\cite{altshuler:09} have argued
that this failure mode dooms the QAA for random instances of NP-complete
problems.  However, the arguments of Altshuler {\it et al.}~have been
criticized by Knysh and Smelyanskiy~\cite{knysh:11}. The application of the
QAA to hard optimization problems has been reviewed recently in
Ref.~\cite{bapst:12}.

Young {\it et al.}~\cite{young:08,young:10} recently examined the performance
of the QAA on random instances of the constraint satisfaction problem called 1-in-3 SAT (to be
described  in the next section) and showed the presence of avoided crossings
associated with very small gaps. These `bottlenecks' appears in a larger and
larger fraction of the instances as the problem size $N$ increases, indicating
the existence of a first order quantum phase transition. This leads to an
exponentially small gap for a \textit{typical} instance, and therefore also to
the failure of adiabatic quantum optimization. 

It is not yet clear to what extent the above behavior found for 1-in-3 SAT is
general and whether it is a feature inherent to the QAA that will plague most
if not all problems fed into the algorithm or something more benign than this.
Previous work~\cite{jorg:08,jorg:10,jorg:10b,hen:11} had argued that a first
order quantum phase transition occurs for a broad class of random optimization
models.

In this paper we contrast the performance of the quantum adiabatic algorithm on random instances of two
combinatorial optimization problems.  The first problem we consider is 3-XORSAT
on a random 3-regular hypergraph, which was studied previously in
Ref.~\cite{jorg:10}.  Interestingly, although this computational problem is
classically easy--an instance can always be solved in polynomial time on a
classical computer by using Gaussian elimination--it is known that classical
algorithms that do not use linear algebra are stymied by this
problem~\cite{franz:01,xor_diff,ricci-tersenghi:11,guidetti:11}.  
In Ref.~\cite{jorg:10} it was shown that the
QAA fails to solve this problem in polynomial time. In this paper we provide
more numerical evidence for this. We also furnish a duality transformation
that helps to understand properties of this model.

The second computational problem we consider is Max-Cut on a 3-regular graph.
This problem is NP-hard. However we consider random instances, for which the computational complexity is
less well understood.

 A nice feature of these problems is that the regularity of the associated hypergraphs
constrains the two ensembles of random instances. Studying the performance of the QAA 
for these problems, we therefore expect to see smaller instance-to-instance 
differences than for the unconstrained ensembles of instances. 

We use two different methods to study the performance of the QAA.  The first
method is quantum Monte Carlo simulation. It is a numerical method that 
is based on sampling paths from the Taylor expansion of the partition function of the system.
Using this method we can extract, for a given instance, the thermodynamic properties
(in particular the ground state energy) as well as 
the eigenvalue gap for the interpolating Hamiltonian $H(s)$. 
This allows us to
investigate the size dependence of the typical minimum gap of the problem from
which we can extrapolate the large-size scaling of the computation time
$\mathcal{T}$ of the QAA. 
The second approach is a quantum cavity method. It is a semi-analytical method that allows us
to compute the thermodynamic properties averaged over
the ensemble of instances in the limit $N\rightarrow \infty$. 
It leads to a set of self-consistent equations that can be solved analytically in some classical 
examples~\cite{cavity,mezard:03b}. However in the quantum case the equations are more complicated and are solved numerically~\cite{laumann:08,krzakala:08}. The method is not exact on general graphs. For locally tree-like random graphs, 
it provides the exact solution of the problem if some assumptions on the Gibbs
measure are satisfied~\cite{cavity,mezard:03b,MM09}.
As we will discuss below the cavity method we use in this paper
gives the exact result for
3-XORSAT, while it only gives an approximation for the Max-Cut problem.  

Using these methods we conclude that the quantum adiabatic algorithm fails to
solve both problems efficiently, although in a qualitatively different way. 

The plan of this paper is as follows. In Section~\ref{sec:models} we describe
the two computational problems that we investigate. In Sec.~\ref{sec:method}
we discuss the methods that we use to obtain our results. These
results are presented in Sec.~\ref{sec:results} and our conclusions are
summarized in Sec.~\ref{sec:conclusions}. Some parts of this paper have
previously appeared in the PhD thesis of one of the authors \cite{gosset11}.

\section{\label{sec:models}Models}

We now discuss in detail the two computational problems 3-regular 3-XORSAT and
3-regular Max-Cut. 

When studying the efficiency of the QAA numerically~\cite{farhi_long:01,young:08,young:10}, it is convenient to
consider instances with a unique satisfying assignment (USA) for reasons that
will be explained in Sec.~\ref{sec:numres}.  On the other hand, the quantum
cavity method is designed to study the ensemble of random instances with no
restrictions on the number of satisfying assignments. In this section we
specify the random ensembles of instances that we investigate in this paper.  

\subsection{\label{sec:XORSAT}3-regular 3-XORSAT}

The 3-XORSAT problem is a clause based constraint satisfaction problem. An
instance of such a constraint satisfaction problem is specified as a list of $M$ logical conditions
(clauses) on a set of $N$ binary variables. The problem is to determine
whether there is an assignment to $N$ bits which satisfies all $M$ clauses.

\begin{figure}
\begin{center}
\includegraphics[width=\figurewidth]{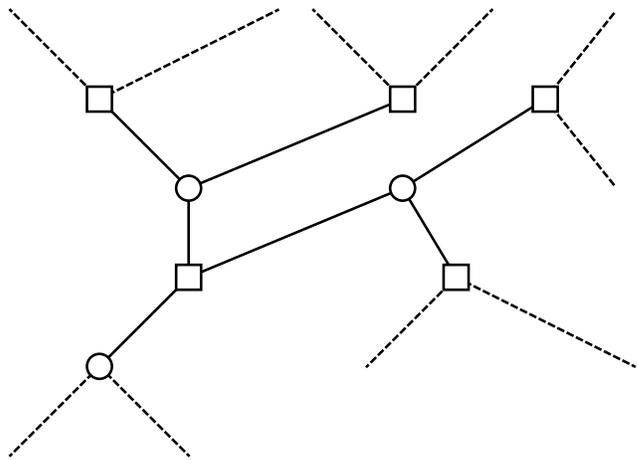}
\caption{
Factor graph of a small part of an instance of the 3-regular 3-XORSAT
problem. In the full factor graph, each clause ($\square$) is connected to
exactly three bits ($\bigcirc$) and each bit is connected to exactly three
clauses, so there are no leaves and the graph closes up on itself. }
\label{fig:3xor2}
\vspace{-0.7cm}
\end{center}
\end{figure}

In the 3-XORSAT problem each clause involves three bits. A given clause is
satisfied if the sum of the three bits (mod 2) is a specified value (either 0
or 1, depending on the clause).  We consider the ``3-regular'' case where
every bit is in exactly three clauses which implies $M= N$. This model has
already been considered by J\"org {\it et al.}~\cite{jorg:10}.  The factor
graph for an instance of 3-regular 3-XORSAT is sketched in
Fig.~\ref{fig:3xor2}.

Since this problem just involves linear constraints (mod 2), the
satisfiability problem can be solved in polynomial time using Gaussian
elimination. However, it is well known that this problem presents difficulties for solvers that do not use linear algebra (see, e.g.\
Refs.~\cite{franz:01,xor_diff,ricci-tersenghi:11,guidetti:11}).  

We associate each instance of 3-regular 3-XORSAT with a problem Hamiltonian
$H_P$ that acts on $N$ spins.  Each clause is mapped to an operator which acts
nontrivially on the spins involved in the clause. The operator for a given
clause has energy zero if the clause is satisfied and energy equal to $1$  if
it is not, so
\begin{equation}
H_{P}=\sum_{c=1}^{N}\left(\frac{1-J_{c}\sigma_{z}^{i_{1},c}\sigma_{z}^{i_{2,}c}\sigma_{z}^{i_{3},c}}{2}\right)\label{eq:3REG3XOR}.
\end{equation}
Here each clause $c\in\{1,...,N\}$ is associated with the 3 bits $i_{1,c},i_{2,c},i_{3,c}$
and a coupling $J_{c}\in\{\pm1\}$ which tells us if the sum of the bits mod 2 should be $0$ or $1$ when the clause is satisfied. 

\subsubsection{Random Instances of 3-regular 3-XORSAT}
As in Ref.~\cite{jorg:10}, we consider both the random ensemble of instances
of this problem and the random ensemble of instances which have a unique
satisfying assignment (USA). In the 3-XORSAT problem as $N\rightarrow\infty$,  instances with a USA are
\textit{a nonzero fraction}, about 0.285~\cite{jorg:10}, of the set of all
instances, so the random ensemble of USA instances should be a good
representation of the fully random ensemble.  

All satisfiable instances (and in particular instances with a USA) 
have the property that the cost function Eq.~(\ref{eq:3REG3XOR})
can be mapped unitarily into the form 
\begin{equation}\label{XOR_ferro}
H_{P}=\sum_{c}\left(\frac{1-\sigma_{z}^{i_{1},c}\sigma_{z}^{i_{2,}c}\sigma_{z}^{i_{3},c}}{2}\right)
\end{equation}
by a product of bit flip operators. 

\subsubsection{Previous Work}

Reference \cite{jorg:10} studied the performance of the QAA on the
random ensemble of instances of 3-regular 3-XORSAT using quantum cavity method
and quantum Monte Carlo simulation.  They also studied the ensemble of random
instances with a USA using exact numerical diagonalization. This work gave
evidence that there is a first order quantum phase transition which occurs at
$s_{c}\approx\frac{1}{2}$ in the ground state. Their results also demonstrate
that the minimum gap is exponentially small as a function of $N$ at the
transition point. 

\subsubsection{Duality Transformation}
\label{duality}

In this section we demonstrate a duality mapping for the ensemble of random
instances of 3-regular 3-XORSAT with a unique satisfying assignment. This
duality mapping explains the critical value $s_{c}=\frac{1}{2}$ of the quantum
phase transition in this model~\cite{jorg:10}.  Consider the Hamiltonian  
\begin{equation}
H(s)=(1-s)\sum_{i=1}^{N}\bigg(\frac{1-\sigma_{x}^{i}}{2}\bigg)+
s\sum_{c=1}^{N}\bigg(\frac{1-\sigma_{z}^{i_{1},c}\sigma_{z}^{i_{2,}c}\sigma_{z}^{i_{3},c}}{2}\bigg)
\label{eq:h_lambda_M} \,.
\end{equation}
Here, the first term is the beginning Hamiltonian and the second term is the
problem Hamiltonian for an instance of 3-regular 3-XORSAT with a unique
satisfying assignment. The 3-regular hypergraph specifying the instance can be
represented by a matrix $M$ where \[ M_{ij}=\begin{cases} 1 & ,\text{ if bit j
is in clause i}\\ 0 & \text{, otherwise. }\end{cases}\] and where $M$ has 3
ones in each row and 3 ones in each column. The fact that there is a unique
satisfying assignment $000...0$ is equivalent to the statement that the matrix
$M$ is invertible over $\mathbb{F}_{2}^{N}$. To see this, consider the
equation (with addition mod 2)
\[
M\vec{v}=\vec{0}.
\] 
This equation has the unique solution $\vec{v}=\vec{0}$ if and only if there
is a unique satisfying assignment for the given instance. This is also the
criterion for the matrix $M$ to be invertible.

The duality that we construct shows that the spectrum of $H(s)$ is the same as
the spectrum of $H_{\text{DUAL}}(1-s)$ where $H_{DUAL}$ is obtained by
replacing the problem Hamiltonian hypergraph by its dual--that is to say, the
instance corresponding to a matrix $M$ is mapped to the instance associated
with $M^T$. The ground state energy per spin (averaged over all 3-regular
instances with a unique satisfying assignment) is symmetric about
$s=\frac{1}{2}$ and the first order phase transition observed in
Ref.~\cite{jorg:10} occurs at $s=\frac{1}{2}$.  For each $c=1,...,N$
define the operator 
\begin{equation} X_{c}=\sigma_{z}^{i_{1},c}\sigma_{z}^{i_{2},c}\sigma_{z}^{i_{3},c}.\label{eq:Xop}\
\end{equation}  
We also define, for each clause $c$, a bit string $\vec{y}^{c}$ \[
\vec{y}^{c}=M^{-1}\hat{e}_{c}.\] Here $\hat{e}_{c}$ is the unit vector with
components $(\hat{e}_{c})_i=\delta_{ic}$. Note that $\vec{y}^{c}$ is the
unique bit string which violates clause $c$ and satisfies all other clauses.
Such a bit string is guaranteed to exist since $M$ is invertible. Let
$y_{i}^{c}$ denote the $i$th bit of the string $\vec{y}^{c}.$ Define, for each
$c=1,...,N$,
\begin{equation} 
Z_{c}=\prod_{i=1}^{N}\left[\sigma_{x}^{i}\right]^{y_{i}^{c}}.\label{eq:zed_c}
\end{equation}  
Note that \[ \{Z_{c},X_{c}\}=0\]  and \[ [Z_{c},X_{c^{\prime}}]=0\text{ for
}c\neq c^{\prime}.\] For each bit $i=1,...,N$ let $c_{1}(i),c_{2}(i),c_{3}(i)$
be the clauses which bit $i$ participates in. Then 
\begin{equation} 
\sigma_{x}^{i}=Z_{c_{1}(i)}Z_{c_{2}(i)}Z_{c_{3}(i)}.\label{eq:xop}
\end{equation}  
This follows from the fact that \[
M\hat{e}_{i}=\hat{e}_{c_{1}(i)}+\hat{e}_{c_{2}(i)}+\hat{e}_{c_{3}(i)}\] and so 
\begin{eqnarray*} 
\hat{e}_{i} & = & M^{-1}\left(\hat{e}_{c_{1}(i)}+
\hat{e}_{c_{2}(i)}+\hat{e}_{c_{3}(i)}\right)\\
& = & \vec{y}^{c_{1}(i)}+\vec{y}^{c_{2}(i)}+\vec{y}^{c_{3}(i)}.
\end{eqnarray*} 
The above equation and the definition Eq.~(\ref{eq:zed_c}) show
Eq.~(\ref{eq:xop}). Now using Eqs.~(\ref{eq:xop}) and~(\ref{eq:Xop}) write 
\begin{eqnarray} 
 H(s)& = & (1-s)\sum_{i=1}^{N}\bigg(\frac{1-\sigma_{x}^{i}}{2}\bigg)\\
&+ &s\sum_{c=1}^{N}\bigg(\frac{1-\sigma_{z}^{i_{1},c}\sigma_{z}^{i_{2,}c}
\sigma_{z}^{i_{3},c}}{2}\bigg)\\
& = & (1-s)\sum_{i=1}^{N}\bigg(\frac{1-Z_{c_{1}(i)}Z_{c_{2}(i)}Z_{c_{3}(i)}}{2}\bigg)\\
&+& s\sum_{c=1}^{N}\bigg(\frac{1-X_{c}}{2}\bigg) \label{final_eq}  \,.
\end{eqnarray} 
The $X$ and $Z$ operators satisfy the same commutation relations as the
operators $\sigma_{x}$ and $\sigma_{z}$. Comparing Eq.~(\ref{final_eq}) with
Eq.~(\ref{eq:h_lambda_M}) we conclude that the spectrum of $H(s)$ is the same
as the spectrum of $H_{DUAL}(1-s)$. This result can be thought of as an
extension of the duality of the one-dimensional random Ising model in a
transverse field, see e.g.\ Ref.~\cite{fisher:95}.

\begin{figure}
\begin{center}
\includegraphics[width=\figurewidth]{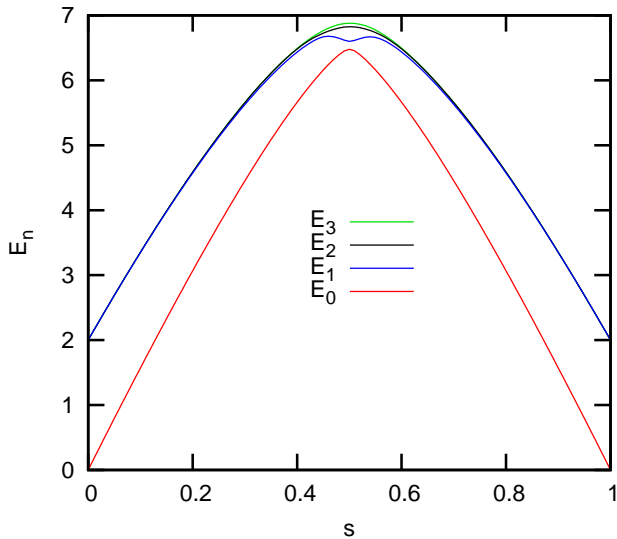}
\caption{(Color online) 
First four energy levels of the interpolating Hamiltonian for a 16-bit
instance of the 3-regular 3-XORSAT problem.  The energy curves for this
instance are close to being symmetric about $s =1/2$. Our duality
transformation means that sending $s\rightarrow(1-s)$  we obtain the spectrum
of the interpolating Hamiltonian for a different instance from the 
same ensemble, obtained by
interchanging the clauses and bits.}
\label{fig:3xor1}
\vspace{-0.7cm}
\end{center}
\end{figure}

In Fig.~\ref{fig:3xor1} we show the first four energy levels of the
interpolating Hamiltonian $H(s)=sH_B+(1-s)H_P$  as a function of $s$ for one
16-bit instance of 3-XORSAT. The duality transformation means that these energy
levels are the same as for the interpolating Hamiltonian
$H'(s)=sH_{P,DUAL}+(1-s)H_B$ which involves the dual instance. Evident from
the figure is the apparent symmetry of the energy levels around $s=1/2$. In
this case the instance and its dual are similar from the point of view of
the QAA. 

The duality argument given here has implications for the phase transition which occurs in the ensemble of random instances of 3-regular 3-XORSAT as $N\rightarrow\infty$. Our numerics in section \ref{sec:results} show that for large $N$, the ground state energy per spin as a function of $s$ (averaged over the ensemble of instances) has a nonzero derivative as $s\rightarrow\frac{1}{2}$. The duality transformation given here implies that this curve is symmetric about $s=\frac{1}{2}$. So there is a discontinuity in the derivative of this curve at $s=\frac{1}{2}$, which is associated with a first order phase transition. 

\subsection{3-regular Max-Cut}

The second model we discuss is also a clause based problem. The instances we
consider are not satisfiable and we are interested in finding the assignment
which gives the maximum number of satisfied clauses. We view this problem as
minimizing a cost function that computes the number of unsatisfied clauses.
The 3-regular Max-Cut problem is defined on $N$ bits, and each bit appears in
exactly three clauses. Each clause involves two bits and
is satisfied if and only if the sum of the two bits (modulo 2) is 1. The
number of clauses is therefore $M=3N/2$. The problem Hamiltonian is

\begin{equation}
H_{P}=\sum_{c}\left(\frac{1+\sigma_{z}^{i_1,c}\sigma_{z}^{i_2,c}}{2}\right) \,.
\label{eq:hamMaxCut}
\end{equation}
The ground state of this Hamiltonian encodes the solution to the Max-Cut
problem.  

The model can also be viewed as an antiferromagnet on a 3-regular random
graph. Because the random graph in general has loops of
odd length, it is not possible to satisfy all of the clauses. 

The Max-Cut problem is NP-hard and accordingly there is no known classical
polynomial time algorithm which computes the ground state energy of the
problem Hamiltonian \eqref{eq:hamMaxCut}. Indeed, even achieving a certain
approximation to the ground state energy is hard, which follows from the fact
that it is NP hard to approximate the Max-Cut of $3$-regular graphs to within
a multiplicative factor $0.997$~\cite{berman:99}.  Interestingly, however,
there is a classical polynomial time algorithm which achieves an approximation
ratio of at least $0.9326$~\cite{halperin:04}.

\subsubsection{Random Instances of 3-regular Max-Cut}

Using the quantum cavity method we study the ensemble of random instances of
3-regular Max-Cut.

The random instances we studied using quantum Monte Carlo simulation were restricted
to those which have exactly 2 minimal energy states (note that this is the
smallest number possible since the problem is symmetric under flipping all the
spins) and for which the ground state energy of the problem Hamiltonian is equal to
$\frac{1}{8}N$.  We choose to study instances with a unique satisfying assignment (up to the bit-flip symmetry of
this problem) because it is numerically more convenient for the extraction of the relevant gap (to
the first even state). For the range of sizes studied, 
$\frac{1}{8}N$ was found
numerically to be the most probable value of the ground state energy.   The restriction to instances with a fixed Max-Cut ($\frac{1}{8}N$) further reduces the
instance-to-instance fluctuations. However, this choice affects the ensemble averaged
value of thermodynamic observables (e.g. the average energy of fully random
instances is different from $N/8$), making it more difficult to compare the quantum Monte Carlo results
with our quantum-cavity results on the fully random ensemble.  We expect (and find
numerically) that this set of instances makes up an exponentially small
fraction of the whole random ensemble for large $N$.

\subsubsection{Previous work}

Laumann et al.~\cite{laumann:08}
used the quantum cavity method to study the transverse field spin glass with
the problem Hamiltonian 
\begin{equation}
H_{P}=\sum_{c}
\left(\frac{1+J_{c}\sigma_{z}^{i_1, c}\sigma_{z}^{i_2, c}}{2}\right)\label{eq:classical_spinglass}
\end{equation}
where each $J_{c}$ is chosen to be $+1$ or $-1$ with equal probability. 

In general there is no {}``gauge transformation'' equivalence between this problem Hamiltonian and the
antiferromagnetic problem Hamiltonian
Eq.~(\ref{eq:hamMaxCut}). However we do expect these models to exhibit
similar properties since a random graph is locally tree-like, and on a tree
such a gauge transformation does exist, see Ref.~\cite{zdeborova:09} for a
discussion of this point in the case where there is no transverse field
present.

Laumann et al. found that this system exhibits a second order phase transition
as a function of the transverse field. Their method is similar to the
quantum cavity method that we use, although the numerics performed in
Ref.~\cite{laumann:08} have some systematic errors which our calculations
avoid. The method used in Ref.~\cite{laumann:08} is a discrete imaginary time
formulation of the quantum cavity method which has nonzero Trotter error,
whereas our calculation works in continuous imaginary time~\cite{krzakala:08}
where this source of error is absent. Our calculation also does not use the
approximation used in Ref.~\cite{laumann:08} where the ``effective action'' of
a path in imaginary time is truncated at second order in a cluster expansion.

\section{\label{sec:method}Method} 

\subsection{\label{sec:numres}Quantum Monte Carlo}

The complexity of the QAA algorithm is determined by
the size dependence of the ``typical'' minimum gap of the problem. Following
Refs.~\cite{young:08, young:10, hen:11}, we analyze the size-dependence of
these gaps by
considering (typically) 50 instances for each size, and then extracting the
minimum gap for each of them.  For each instance, we perform quantum Monte
Carlo simulations for a range of $s$ values and hunt for the minimum gap. We
then take the median value of the minimum gap among the different instances
for a given size to obtain the ``typical'' minimum gap. In situations where
the distribution of minimum gaps is very broad, the average can be dominated
by rare instances which have a much bigger gap than the typical one, and so
the typical value (characterized, for example, by the median) is a better
measure than the 
average. This has been discussed in detail by Fisher~\cite{fisher:95} who
solved the random transverse field model in one-dimension exactly, and found
that the average gap at the quantum critical point vanishes polynomially with
$N$ while the typical gap has stretched exponential behavior,
$\exp(- c N^{1/2})$.

Quantum Monte Carlo simulation works by sampling random variables from a probability distribution (over some configuration space) which contains information about the quantum system of interest. The probability distribution is sampled by Markov chain Monte Carlo, and properties of the quantum system to be studied are obtained as expectation values. Different quantum Monte Carlo methods are based on different ways of associating probability distributions to a quantum system.

In our simulations we use a quantum Monte Carlo technique known as the stochastic series expansion (SSE) algorithm~\cite{sandvik:99,sandvik:92}.  In this method, the probability distribution associated with the quantum system is derived from the Taylor series expansion of the partition function $\Tr[\e^{-\beta \hat{H}}]$ at inverse temperature $\beta$. This is in contrast to other quantum Monte Carlo techniques which are based on the path integral expansion of the partition function.  Whereas some of these techniques have systematic errors because the path integral expansions used are inexact, the SSE that we use has no such systematic error.  

A second feature of the SSE method that we use is that the Markov chain used
to sample configurations allows global, as well as local, updates, 
which leads to faster equilibration. We further speed up equilibration by
implementing ``parallel tempering''~\cite{hukushima:96}, where simulations for
different values of $s$ are run in parallel and spin configurations with
adjacent values of $s$ are swapped with a probability satisfying the detailed
balance condition.  Traditionally, parallel tempering is performed for systems
at different temperatures, but here the parameter $s$ plays the role of
(inverse) temperature~\cite{sengupta:02}.

The details of our implementations of the SSE method for 3-XORSAT and Max-Cut are slightly different and are further discussed 
in Appendix~\ref{app:details_QMC}.
Moreover, the Max-Cut problem can not be simulated with the `traditional' SSE method because of its symmetry under flipping all of the spins:  
For a given instance of Max-Cut, every eigenstate of the interpolating Hamiltonian $H(s)$ is either even or odd under this symmetry transformation. Since the ground state is even, here we are interested in the eigenvalue gap to the first even excited state. We therefore design our quantum Monte Carlo simulation so that it works in the subspace of even states.
The modified algorithm is detailed in Appendix~\ref{app:projSSE}.

In our simulations we extract the gap from imaginary time-dependent correlation functions.
The gap of the system for a given instance and a given $s$ value is extracted 
by analyzing measurements of (imaginary) time-dependent correlation functions
of the type
\begin{equation}
C_A(\tau) = \langle \hat{A}(\tau) \hat{A}(0) \rangle -\langle A \rangle^2  \,,
\end{equation}
where the operator $\hat{A}$ is some measurable physical quantity. 
It is useful to optimize the choice of correlation functions such that
the contribution from the first excited state, $m=1$ in Eq.~\eqref{Ctau} below, is as
large as possible relative to the contributions from higher excited
states. One way of doing this, which was used in some of the runs,
is described in Ref.~\cite{hen:12}.

The evaluation of $\langle A \rangle^2$ in the above equation is computed from
the product $\langle A \rangle^{(1)} \langle A \rangle^{(2)}$ where the two
indices correspond to different independent simulations of the same system.
This eliminates the bias stemming from straightforward squaring of the
expectation value. 

In the low temperature limit, $T \ll \Delta E_1$ where $\Delta E_1 = E_1 - E_0$,
the system is in its ground state so
the imaginary-time correlation function is given by
\begin{equation}
C_A(\tau) = \sum_{m=1} |\langle 0 | \hat{A} | m \rangle|^2
\left( e^{-\Delta E_m \tau} + e^{-\Delta E_m (\beta - \tau)} \right)  \,,
\label{Ctau}
\end{equation}
where $\Delta E_m = E_m - E_0$. At 
long times, $\tau$, the correlation function is dominated by the
smallest gap $\Delta E_1$
(as long as the
matrix element $|\langle 0 | \hat{A} | 1 \rangle|^2$ is nonzero).  On a log-linear plot
$C_A(\tau)$, then has a region where it is a straight line whose slope
is the negative of the gap. This can therefore be extracted by 
linear fitting.
A  more detailed description of the method may be found in Ref.~\cite{hen:12}. 

\subsection{The quantum cavity method}

The quantum cavity method~\cite{krzakala:08,laumann:08} is a technique that
is used to study thermodynamic properties of transverse field spin
Hamiltonians. In our implementation we use the continuous imaginary time method
from Ref.~\cite{krzakala:08}. Quantum Cavity methods have now been used to study a number of problems including the
ferromagnet on the Bethe lattice in uniform~\cite{krzakala:08} and random~\cite{dimitrova:2011} transverse field, 
the spin glass on
the Bethe Lattice~\cite{laumann:08}, 3-regular 3-XORSAT~\cite{jorg:10}, and the quantum Biroli-M\'ezard model~\cite{2011PhRvB..83i4513F}. 

If the Hamiltonian is a two-local transverse field Hamiltonian on a finite
number of spins and if the interaction graph consists of a tree (i.e if there
are no loops) then the quantum cavity equations are exact. In this case the
quantum cavity equations are a closed set of equations that exactly
characterize the thermodynamic properties of the system at a fixed inverse
temperature $\beta$. If instead the interaction graph is a random regular
graph with a finite number of spins then it must have loops. As $N\rightarrow \infty$ we can think of it as an ``infinite tree'' since the typical
size of loops in such a graph diverges.

Quantum
cavity methods (such as the one we use) for problems defined on random
regular graphs make use of two properties of the system:
{\it (i)} the fact that a random
regular graph is locally tree-like;
{\it (ii)} the fact that spin-spin correlations decay quickly as a function of distance.
While the first property is true with probability 1 for random regular graphs when $N\to\infty$,
the second property is not always true and we now discuss it more carefully.

The simplest case is when the Gibbs measure is characterized by a single pure state that has
the clustering property, as in a paramagnetic phase. This happens at high enough temperature
or large enough transverse field. In this case, correlations decay exponentially and the simplest version
of the cavity method (the so-called ``replica symmetric (RS)'' cavity method) gives the exact result.
Upon lowering the temperature or the transverse field, a phase transition towards a more complicated
phase can be encountered. If this phase is a standard broken-symmetry phase (e.g. a ferromagnetic
phase), then correlation decay holds provided one adds an infinitesimal symmetry-breaking field,
and the RS cavity method still provides the exact result~\cite{krzakala:08}.

However, if the transition is to a spin glass phase, then the Gibbs measure is
split into a large number of states, and the decorrelation property that is
required by the cavity method only holds within each state. In this case,
there is no explicit symmetry breaking, therefore the states cannot be
selected by adding an infinitesimal external field.  It turns out that refined
versions of the cavity method must be used, that are based on assumptions on
the structure of these states~\cite{MM09}.  The simplest assumption is that
states are distributed in a uniform way in the phase space of the system, and
leads to the so-called ``1 step replica symmetry breaking (1RSB)''
approximation.  In more complicated cases, states might be arranged in
``clusters'' leading to a hierarchical organization, and this requires further
steps of RSB~\cite{mezard:87}. A consistency check can be performed within the
method to check whether a given RSB scheme gives the exact result or whether
further RSB steps are required.

For the XORSAT problem on random regular graphs, it can be rigorously shown
that the 1RSB scheme gives the exact result in the classical
case~\cite{xor_1,xor_2}, and it has been conjectured that the same is true for
the quantum problem in transverse field~\cite{jorg:10}.  For the specific case
of 3-regular 3-XORSAT investigated here, a RS calculation is enough to get the
thermodynamic properties~\cite{jorg:10}, which is why we use the RS method
in our simulations of this model. We study this problem at a temperature low enough that no residual
temperature dependence of the energy is observed. 
Furthermore, we set the parameters of our calculation to be more computationally demanding than those of reference~\cite{jorg:10}, which allows us to achieve better precision.

The study of 3-regular Max-Cut is more involved. To understand how well the cavity method works on this problem, we can look at results obtained for the classical 3 regular spin-glass with Hamiltonian \eqref{eq:classical_spinglass}. We expect that these problems have very similar (possibly identical) thermodynamic properties~\cite{zdeborova:09}. For the classical 3-regular spin glass it can be shown that neither the RS nor the 1RSB cavity method give exact results~\cite{cavity,mezard:03b,zdeborova:09}, and it is widely believed that
an infinite number of RSB steps is required (this has been shown rigorously
for the $z$-regular spin glass as $z\to\infty$, which corresponds to the
Sherrington-Kirkpatrick model~\cite{mezard:87}).  However the 1RSB calculation
gives a good approximation to the classical ground state
energy~\cite{mezard:03b}. 

To study 3-regular Max-Cut we used the 1RSB quantum cavity method with Parisi
parameter $m=0$. This level of approximation is more accurate than the replica symmetric approximation but less accurate than the full 1RSB calculation. With this method we studied the $N\rightarrow\infty$ limit of the random ensemble of 3-regular Max-Cut Hamiltonians 
\[
H(\lambda)=\sum_{c}\sigma_z^{i_1,c}\sigma_z^{i_2,c}-\lambda\sum_{i=1}^{N}\sigma_x^{i}.
\]

We ran our calculations at finite inverse temperature
$\beta=4$ and various values of the transverse field $\lambda$.  Thermodynamic
expectation values with respect to $H(\lambda)$ at inverse temperature $\beta$
can be related to thermodynamic expectation values with
respect to $H(s)=(1-s)H_B+sH_P$ at $s=\frac{1}{1+\lambda}$ and inverse temperature
$\beta'=\frac{2\beta}{s}$.  Note that there is an $s$ dependence introduced in
the temperature when relating these two thermodynamic ensembles.

\section{\label{sec:results}Results}

In what follows we present the results of the QMC simulation alongside those
of the quantum cavity results for the two problems we study here. We show that
the QAA fails with the choice of interpolating Hamiltonians discussed
previously; for both problems the running time appears to be exponentially long as a
function of the problem size. However, the reasons for this failure are
different for each of the models.

\subsection{Random 3-regular 3-XORSAT}
The 3-regular 3-XORSAT problem was studied by J\"org {\it et
al.}~\cite{jorg:10} who determined the minimum gap for sizes up to $N= 24$.
Here, we extend the range
of sizes up to $N= 40$ by quantum Monte Carlo simulations. The two sets of
results agree and provide compelling evidence for an exponential minimum gap.
The duality argument in Sec.~\ref{duality}, shows that the quantum phase
transition occurs exactly at $s = s_c = 1/2$. Our numerics show that the phase transition is strongly first
order, in agreement with Ref.~\cite{jorg:10}.

\begin{figure}
\begin{center}
\includegraphics[width=\figurewidth]{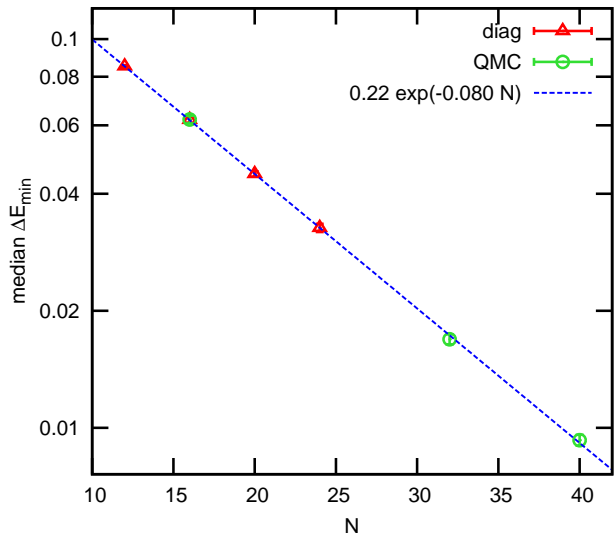}
\caption{(Color online)
Median minimum gap as a function of problem size of the 3-regular 3-XORSAT
problem on a log-linear scale. The straight-line fit is good, indicating
an exponential dependence which in turn leads to an exponential complexity of
the QAA for this problem.  Triangles indicate exact-diagonalization results
while the circles are the results of QMC simulations.}
\label{fig:3xorDE}
\vspace{-0.7cm}
\end{center}
\end{figure}

We show results for the median minimum gap as a function of size for the
3-regular 3-XORSAT problem in Fig.~\ref{fig:3xorDE} (log-lin).  A straight line
fit works well for the log-lin plot, which provides evidence that
the minimum gap is exponentially small in the system size. The results shown here generalize and
agree with those obtained by J\"org {\it et al.}~\cite{jorg:10}. While
Ref.~\cite{jorg:10} computed the average
minimum gap and we computed the median, the
difference here is very small because the distribution of minimum gaps is
narrow for this problem, see for example Fig.\ 21 of Ref.~\cite{bapst:12}. 

\begin{figure}
\begin{center}
\includegraphics[width=\figurewidth]{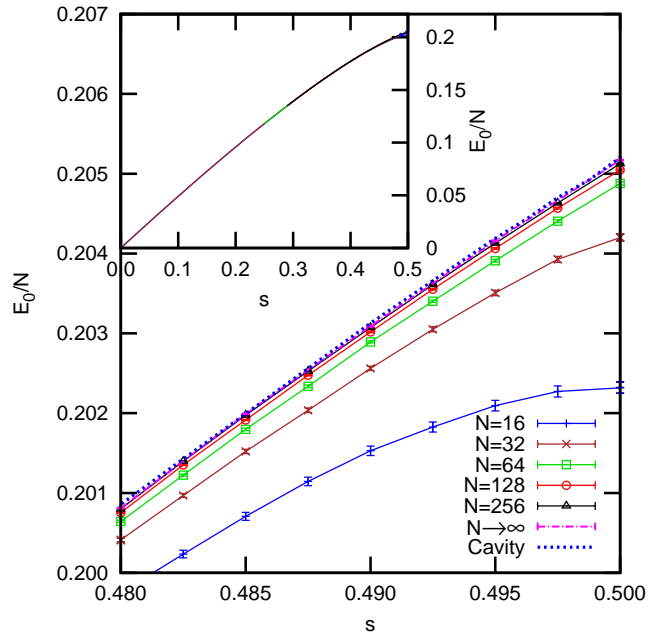}
\caption{(Color online)
Mean energy (averaged over 50 sample instances per size) of the 3-regular
3-XORSAT problem as a function of the adiabatic parameter $s$ for different
sizes (QMC results) compared with the RS quantum cavity
calculations. Because of the duality of the model, the true curve (averaged
over all instances at a given value of $N$) is symmetric about $s=1/2$. The
main panel shows a blowup near the symmetry point $s=1/2$.  In the inset, the
entire range is shown.}
\label{fig:E03XOR}
\vspace{-0.7cm}
\end{center}
\end{figure}

\begin{figure}
\begin{center}
\includegraphics[width=\figurewidth]{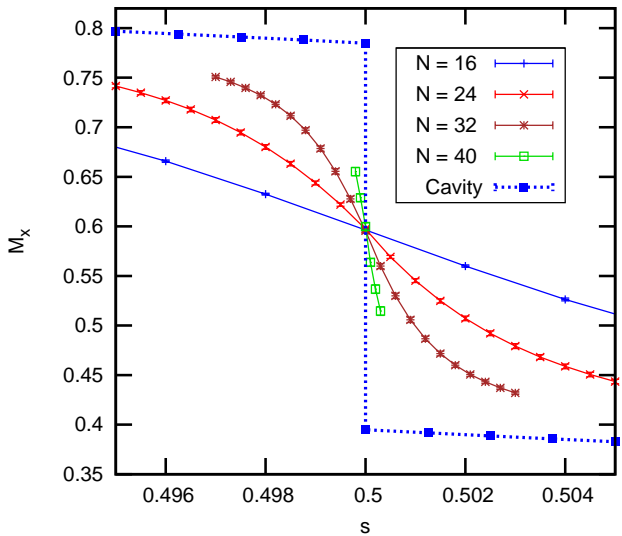}
\caption{(Color online) Magnetization along the $x$-axis,
$M_x = N^{-1} \sum_i \langle \sigma^x_i \rangle$, as a function of the
adiabatic parameter $s$ for the 3-regular 3-XORSAT problem. Results obtained both by QMC and the cavity method are shown. The latter
indicates a sharp discontinuity at $s = s_c = 1/2$. 
The slope of the QMC results at $s=1/2$ increases with increasing $N$,
consistent with a discontinuity at $N = \infty$.}

\label{fig:Sx3XOR}
\vspace{-0.7cm}
\end{center}
\end{figure}

\begin{figure}
\begin{center}
\includegraphics[width=\figurewidth]{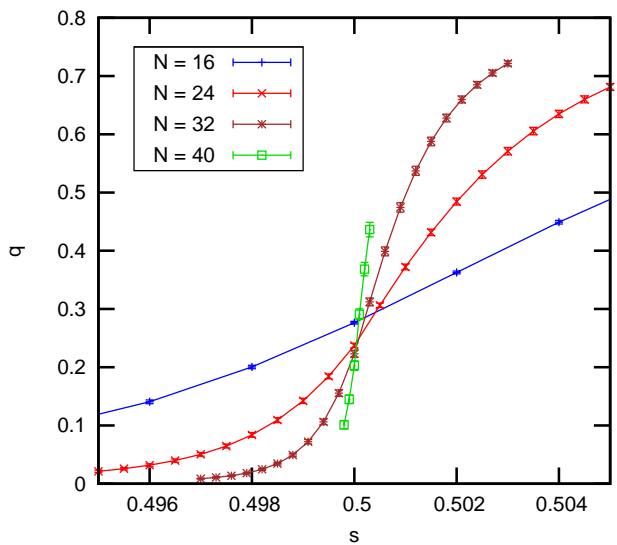}
\caption{(Color online)
The spin-glass order parameter $q$ as defined in Eq.~\eqref{eq:q} as a function
of the adiabatic parameter $s$ for the 3-regular 3-XORSAT problem.  The rapid
change for large sizes around $s=1/2$
indicates a first-order quantum phase transition at this value of $s$.}
\label{fig:q3XOR}
\vspace{-0.7cm}
\end{center}
\end{figure}

We also computed some ground-state properties of the model: the energy
$\langle \hat{H} \rangle$, the magnetization along the $x$-axis $M_x=\frac1{N}
\sum_i \langle \sigma_i^x \rangle$, and the spin-glass order parameter defined
by:
\begin{equation}
\label{eq:q}
q= \frac1{N} \sum_i \langle \sigma_i^z \rangle^2 \,.
\end{equation}
These quantities, averaged over 50 instances for each size,
are plotted in Figs.~\ref{fig:E03XOR},
\ref{fig:Sx3XOR}, and~\ref{fig:q3XOR}.

\begin{figure}
\begin{center}
\includegraphics[width=\figurewidth]{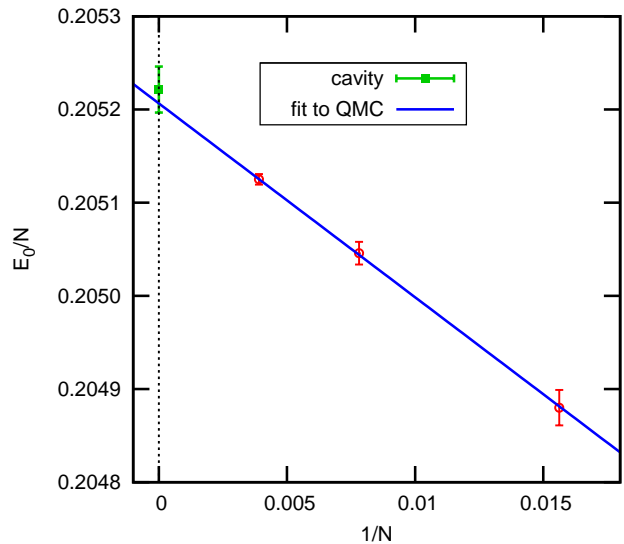}
\caption{(Color online)
Extrapolation of the energy values as given by the QMC method (solid line) for
different system sizes at $s=1/2$ as compared to the value given by the cavity
method (which is for $N = \infty$) for the 3-regular 3-XORSAT problem, assuming a $1/N$ dependence. Extrapolating the QMC results to $N = \infty$
seems to given a result consistent with the cavity value.
}
\label{fig:extrap_en3XOR}
\vspace{-0.7cm}
\end{center}
\end{figure}

Figure \ref{fig:E03XOR} shows that for large system sizes differences between the QMC results for the
ground state energy and the (replica symmetric) cavity results
are small. Reference \cite{jorg:10} has argued that the replica symmetric
(RS) cavity method is actually exact for the thermodynamic properties of the 3
regular 3-XORSAT problem. To check this, in Fig.~\ref{fig:extrap_en3XOR} we
have expanded the vertical scale and show an extrapolation of the QMC results
to $N=\infty$ at the critical value $s=s_c=1/2$ (where finite-size corrections
are largest).  The extrapolated value appears to be consistent
with the cavity result.

The rapid variation of $M_x$ and $q$ shown in figures \ref{fig:Sx3XOR}
and~\ref{fig:q3XOR} in the vicinity of $s_c = 1/2$ is evidence for a
first-order transition.  Figure
\ref{fig:Sx3XOR} also shows a discontinuity in the $x$-axis
magnetization predicted by the cavity calculations at $s=\frac{1}{2}$. In the quantum Monte Carlo data we see that the slope of the
magnetization increases with $N$ and is therefore consistent with the cavity
prediction for $N\rightarrow \infty$.

\subsection{Random 3-regular Max-Cut}

In the Monte Carlo simulations of the Max-Cut problem we
restrict ourselves to instances for which the problem Hamiltonian has a ground
state degeneracy of two, and for which the ground state energy is $N/8$. For this
ensemble of instances, we measured the
energy, the $x$-magnetization and the spin-glass order parameter using quantum
Monte Carlo simulations. Because of the bit-flip symmetry of the
model, we use the following different definition of the spin glass order
parameter:
\begin{equation}
\label{eq:q2}
q'= \left({1\over N(N-1)}\,  \sum_{i\ne j} \langle \sigma_i^z \sigma_j^z
\rangle^2 \right)^{1/2}\,. 
\end{equation}

\begin{figure}[H]
\begin{center}
\includegraphics[width=\figurewidth]{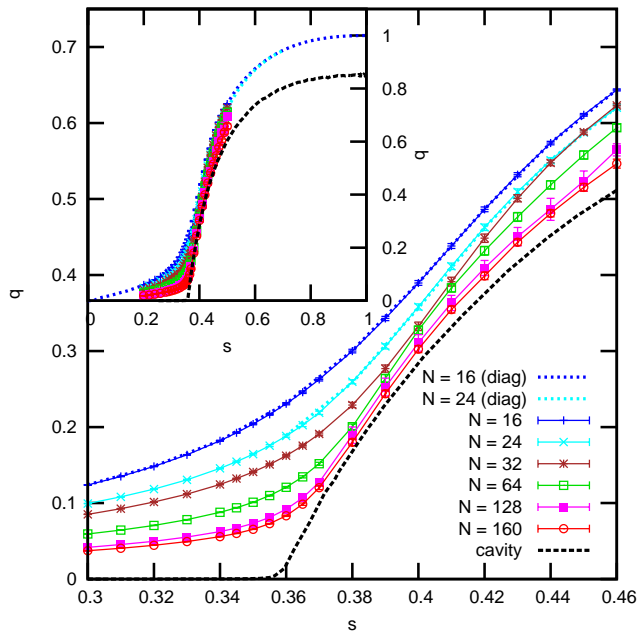}
\caption{(Color online) 
The spin-glass order parameter $q'$ obtained from Monte Carlo
simulations, obtained from Eq.~(\ref{eq:q2}), as
a function of the adiabatic parameter $s$ for the Max-Cut problem. Also shown
is the value of $\bar{q}$ from the cavity calculation, which is defined differently as discussed in the text. The inset shows a
global view over the whole range of s, indicating large differences between
the Monte Carlo and cavity calculations for large $s$. This may be due to the
different ensembles used in the two calculations, as discussed in the text.
}
\label{fig:qMax}
\end{center}
\end{figure}

\begin{figure}[H]
\begin{center}
\includegraphics[width=\figurewidth]{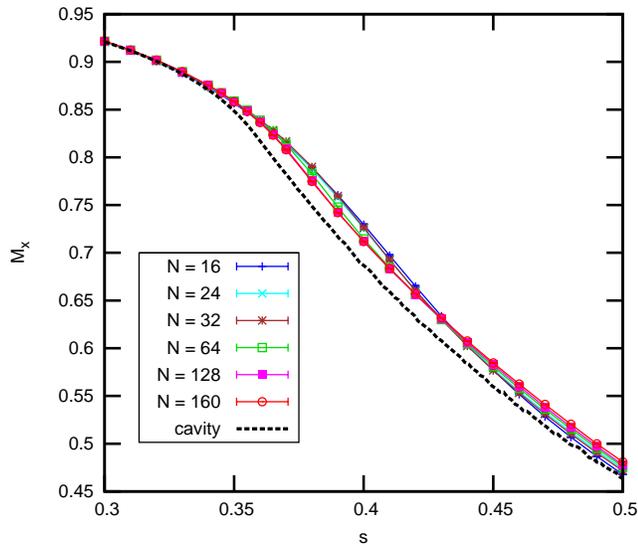}
\caption{(Color online)
Magnetization along the $x$ direction, $M_x = N^{-1}
\sum_i \langle \sigma^x_i \rangle$, as a function of the adiabatic
parameter $s$ for the Max-Cut problem.}
\label{fig:SxMax}
\end{center}
\end{figure}

\begin{figure}[H]
\begin{center}
\includegraphics[width=\figurewidth]{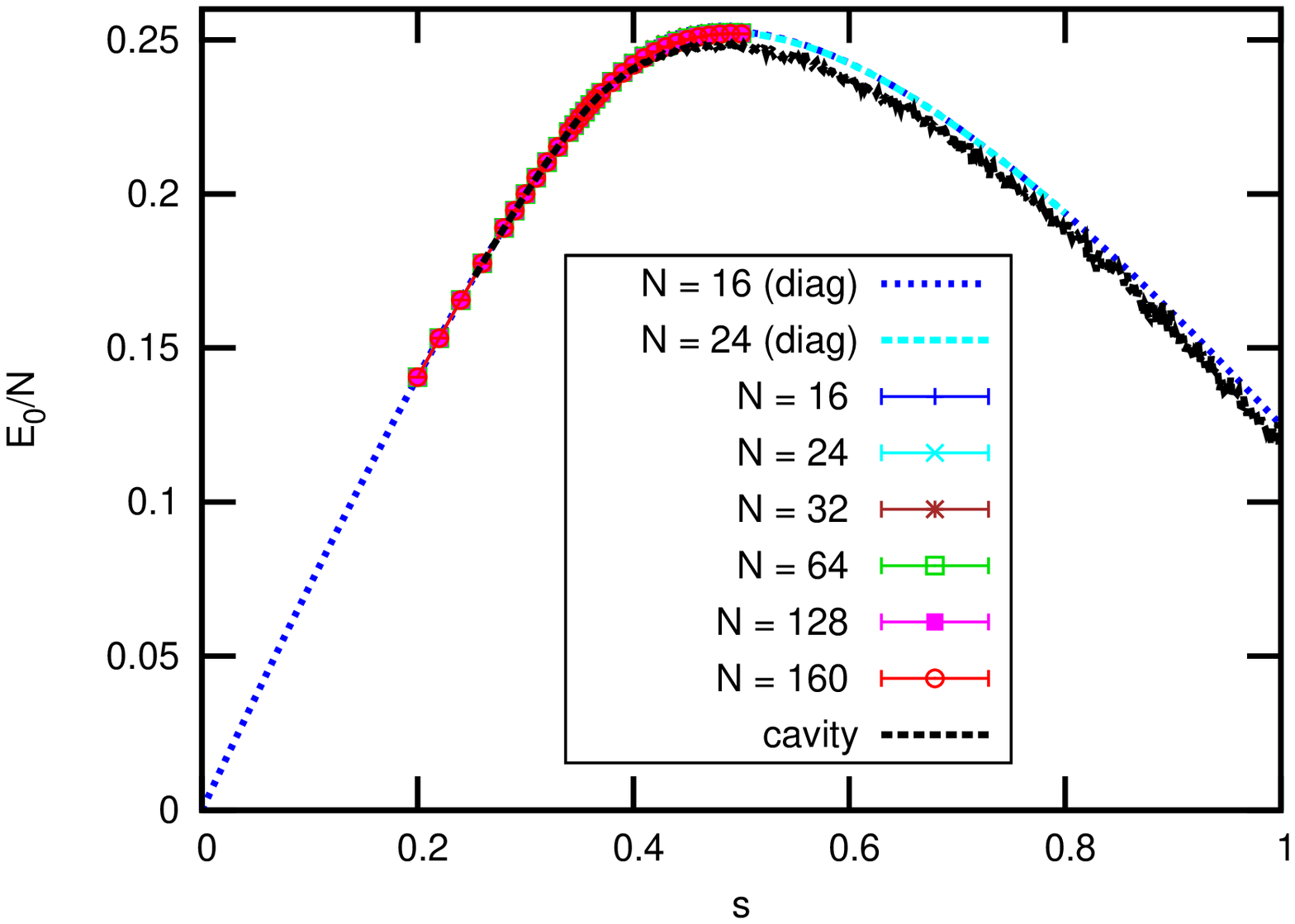}
\includegraphics[width=\figurewidth]{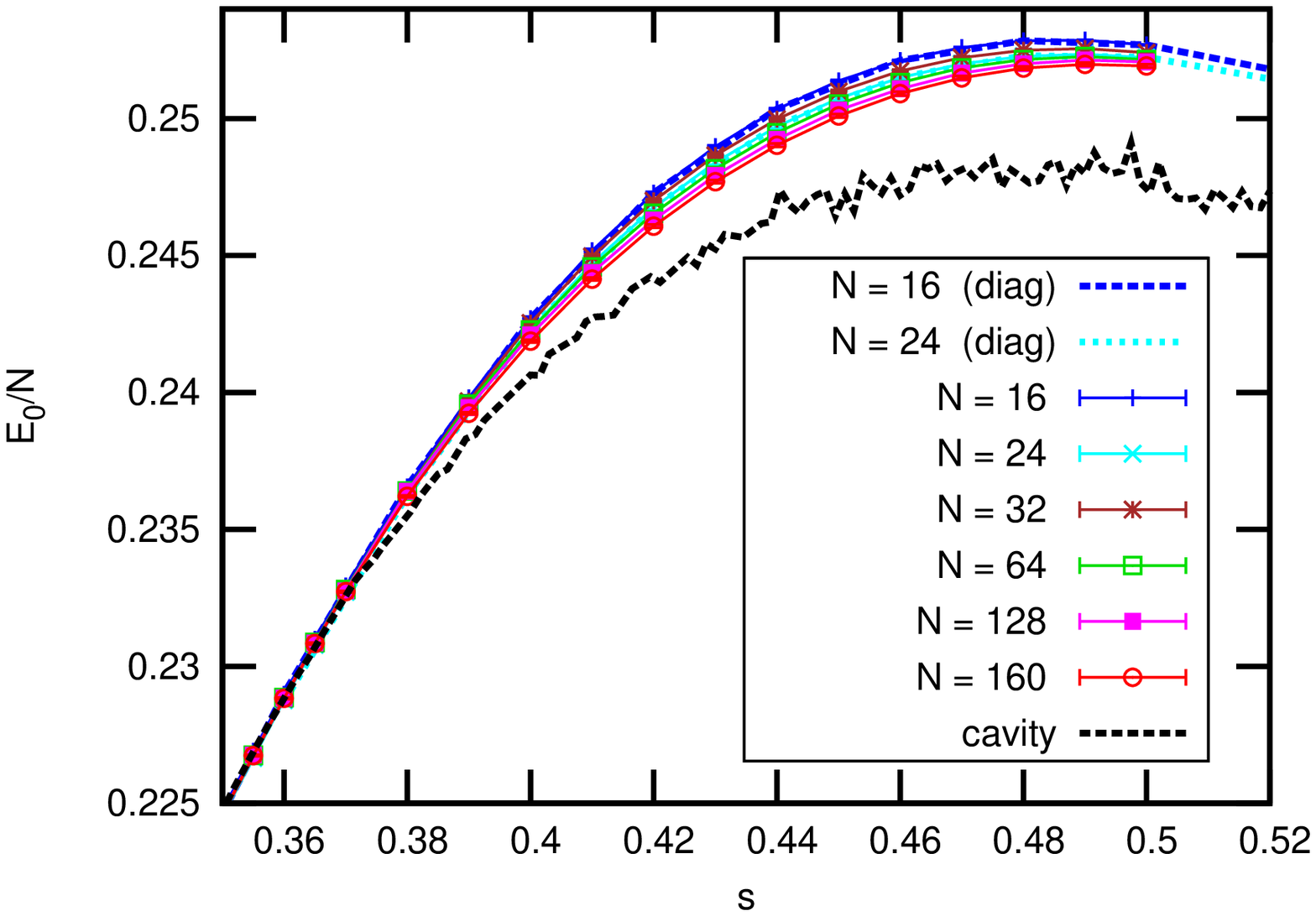}
\caption{(Color Online) Energy as a function of the adiabatic parameter $s$
for the Max-Cut problem. The cavity results computed at inverse temperature $
\beta=\frac{8}{s}$ are depicted by the dashed line. The lower panel is a blow up of the
region around the maximum, which illustrates the difference between the Monte
Carlo and cavity results. Some of this
difference may be due
to the different ensembles used, as discussed in the text.
}
\label{fig:E0Max}
\end{center}
\end{figure}

In Figs.~\ref{fig:qMax}, \ref{fig:SxMax} and \ref{fig:E0Max} we compare the QMC results with those of our quantum cavity method computation.  Recall that the cavity method
results apply to the random ensemble of instances. Formally the value of the spin glass parameter $q$ from Eq.~\eqref{eq:q} is zero due to the bit flip symmetry of the Hamiltonian. However, the cavity method works in the thermodynamic limit in which this symmetry is spontaneously broken for $s$ greater than the critical value $s_c$. For the cavity calculations, we measure a spin glass order parameter $\bar{q}$ which is the
magnetization squared for each thermodynamic ``state'', averaged over the
states. This becomes non-zero for $s > s_c \simeq 0.36$ as shown in
Fig.~\ref{fig:qMax}.  The Monte Carlo results are consistent with this value
of $s_c$. The inset of Fig.~\ref{fig:qMax} shows a global view of the two
spin glass order parameters that we measure, over the whole range of $s$.
There we see differences between the (different) order parameters measured
using Monte Carlo and cavity calculations for large $s$. Note that the Monte
Carlo simulations take only instances with a doubly degenerate ground state
for the problem Hamiltonian, $s=1$, so $q'=1$ in this limit, whereas the cavity
calculations are done for the random ensemble where the instances have much
larger degeneracy at $s=1$ and so $\bar{q} < 1$ in this limit.

Our numerical results for the $x$-component of the magnetization are shown in 
Fig.~\ref{fig:SxMax}. We see no evidence of a discontinuity in this quantity at $s_c \simeq 0.36$ for large $N$. This is in contrast with the corresponding plot for 3-XORSAT in figure \ref{fig:Sx3XOR}.

Results for the energy of the Max-Cut problem, obtained both from
Monte Carlo and the cavity approach are shown in Fig.~\ref{fig:E0Max}. The two
agree reasonably well but there are differences in the spin glass phase, $s >
s_c$, which may be due to the different ensembles used in the two
calculations. We also note that our cavity method
computation is performed at nonzero temperature.

\begin{figure}
\begin{center}
\includegraphics[width=\figurewidth]{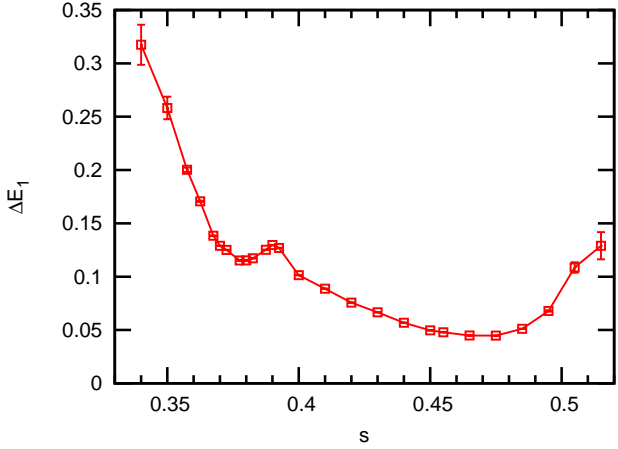}
\caption{(Color online) The gap to the first (even) excited state as a function of the adiabatic
parameter $s$ for one of the $N=128$ instances of the Max-Cut problem,
showing two distinct minima.
The first, higher, minimum is close to $s \approx 0.36$ (the location of the
order-disorder phase transition) while the other, lower minimum
(global in the range)
is well within the spin-glass phase.}
\label{fig:128gap}
\vspace{-0.7cm}
\end{center}
\end{figure}

Using quantum Monte Carlo simulations we have determined the energy gap as a function
of $s$ for $s$ in the range between $0.3$ (i.e.\ well below $s_c$) and $0.5$
(i.e.\ well above $s_c$) for sizes between $N = 16$ and $160$. For the smaller
sizes we find a single minimum in this range, which lies a little above $s_c$.
However, for larger sizes, we see a fraction of instances in which there is a
minimum close to $s_c$ and a second, deeper, minimum for $s > s_c$ well inside the spin-glass phase. A set
of data which shows two minima is presented in Fig.~\ref{fig:128gap}.
This interesting behavior of the minimum gaps suggests the following
interpretation: The minima found close (just above) $s_c$ correspond to the
order-disorder quantum phase transition. Above $s_c$ the system is the
spin-glass phase. The minima that are well within the spin-glass phase may
correspond to `accidental' or perturbative crossings in the spin glass. 

Double-minima occurrences become more frequent as
the system size increases.  While no double minima were found for
sizes $N=16, 24$ and $32$ (within the studied range of $s$ values), for sizes
$N=64, 128$ and $160$ the percentage of instances that exhibit such double
minima was found to be approximately $7\%$, $36\%$ and $40\%$, respectively
(obtained from $\sim 50$ instances for each size).

\begin{figure}
\begin{center}
\includegraphics[width=\figurewidth]{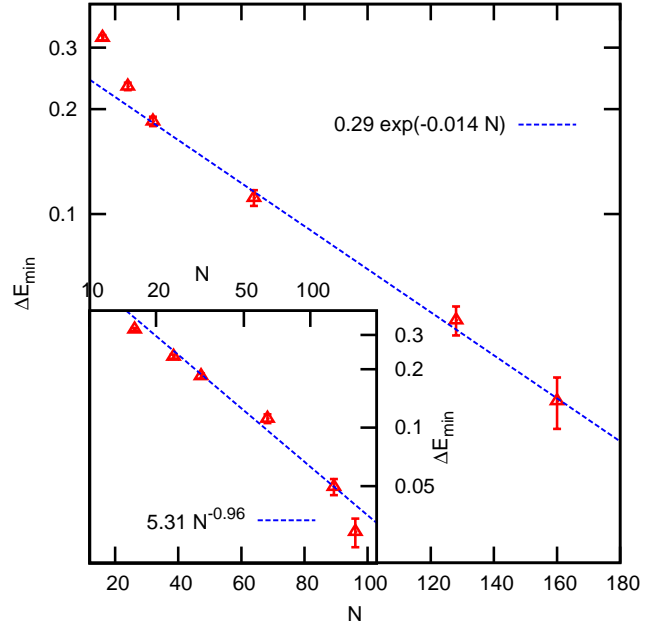}
\caption{(Color online)
Median minimum gap, on log-linear (main panel) and log-log (inset) scales, for
the 3-regular Max-Cut problem for $s$ in the range 0.3 to 0.5. 
The straight-line fit on the log-linear scale (omitting the two smallest
sizes)
is a much better fit ($Q=0.57$) than that of the log-log
scale ($Q=2.7 \times 10^{-3}$), in which the two smallest sizes are also
omitted.}
\label{fig:maxcutDE}
\end{center}
\end{figure}

We have therefore performed two analyses on the data for the gap. In the first
analysis we determine the global minimum (for the range of $s$ studied) for
each instance and determine the median over the instances. There are about 50
instances for each size. This data is presented in Table~\ref{table:gap1}, and is plotted in Fig.~\ref{fig:maxcutDE} 
both as log-lin (main figure) and log-log (inset). A straight line fit works
well for the log-lin plot (goodness of fit parameter $Q = 0.57$), provided
that we omit the two \textit{smallest}
sizes. The goodness of fit parameter $Q$ is the probability that, given the fit,
the data could have the observed value of $\chi^2$ or greater,
see Ref.~\cite{press:92}.
However, a straight-line fit
works much less well for the
log-log plot ($Q = 2.7 \times 10^{-3}$), again omitting the two smallest
sizes, because the data for the
\textit{largest} size lies below the extrapolation from smaller sizes. 
If smaller points do not lie on the fit, it is possible that the fit is correct and the 
deviations are due to corrections to scaling. However, if the largest size
shows a clear deviation then the fit can not describe the asymptotic
large-$N$ behavior. From these fits we conclude that an exponentially decreasing gap is
preferred over a polynomial gap.

\begin{table}[H]
\caption{Median minimum gap for 3-regular Max-Cut (plotted in figure \ref{fig:maxcutDE})} 
\centering 
\begin{tabular}{c c c c} 
\hline\hline 
N & Median gap & Error \\ [0.5ex] 
\hline 
 16   & 0.3203   & 0.0056 \\
 24   &  0.2323   & 0.0057\\
 32    & 0.1844    & 0.0057\\
 64    & 0.1113    & 0.0058\\
128   &  0.0496    & 0.00473\\
160    & 0.0291   & 0.0049\\ [1ex]
\hline
\end{tabular}
\label{table:gap1} 
\end{table}

\begin{figure}
\begin{center}
\includegraphics[width=\figurewidth]{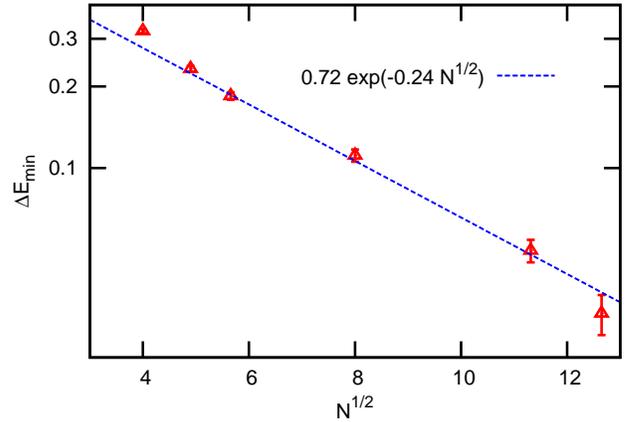}
\caption{(Color online) A stretched exponential fit to
$A e^{-c N^{1/2}}$ for data for the median minimum gap for the 3-regular Max-Cut 
problem, omitting the two smallest sizes. The fit is satisfactory ($Q = 0.31$).
}
\label{fig:maxcutDE_str_exp}
\vspace{-0.7cm}
\end{center}
\end{figure}
\begin{figure}
\begin{center}
\includegraphics[width=\figurewidth]{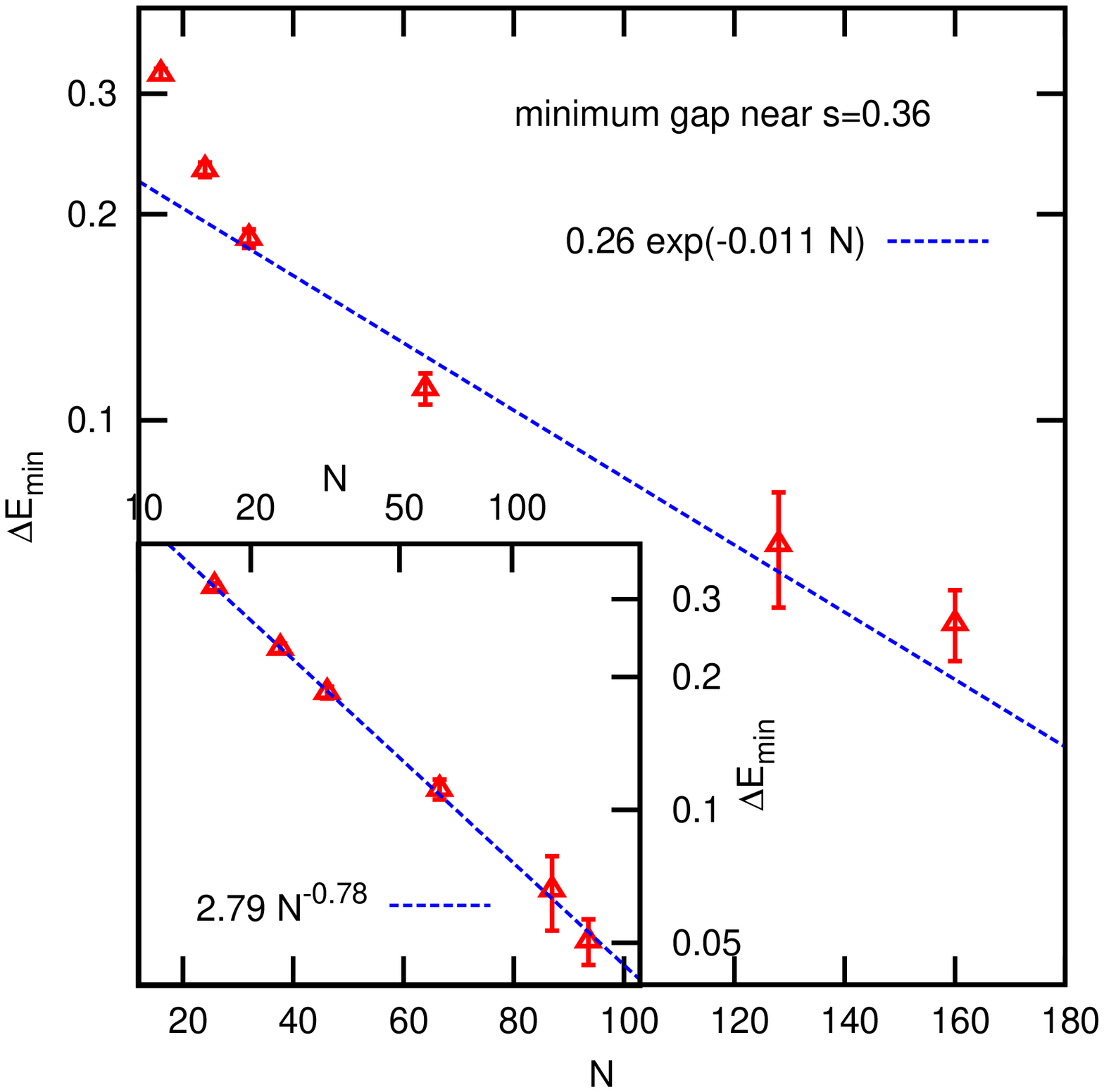}
\caption{(Color online) Median minimum gap, on log-linear (main panel) and
log-log (inset) scales, as a function of problem size for the 3-regular Max-Cut
problem. Here, the minimum gaps were taken from the vicinity of the quantum
phase transition at $s = s_c \simeq 0.36$ and are therefore not necessarily the
global minima. The two fits indicate that in this case the polynomial
dependence is more probable.}
\label{fig:maxcutLocalDE}
\vspace{-0.7cm}
\end{center}
\end{figure}

There are other possibilities for the scaling of the minimum
gap with size, in addition to polynomial or
exponential. For example, we considered a 
``stretched exponential''  scaling of the form $A e^{-c N^{0.5}}$.  Omitting the first two
points (as we did for the exponential fit) we find the fit is satisfactory, as shown in the upper panel of
Fig.~\ref{fig:maxcutDE_str_exp} ($Q = 0.31$). Hence it is possible that the minimum gap decreases as a stretched exponential.

However, if for the instances with more than one minimum, we just take the
minimum close to the critical value $s_c$, a different picture emerges, as shown in
Fig.~\ref{fig:maxcutLocalDE}. In this case, a
straight line fit works well for
the log-log plot ($Q=0.96$), but poorly for the log-lin plot
($Q = 0.016$).  For consistency, we again omitted
the two smallest sizes for the log-lin plot. These
results indicate that the gap
only decreases polynomially with size near the quantum critical point.

\begin{figure}
\begin{center}
\includegraphics[width=\figurewidth]{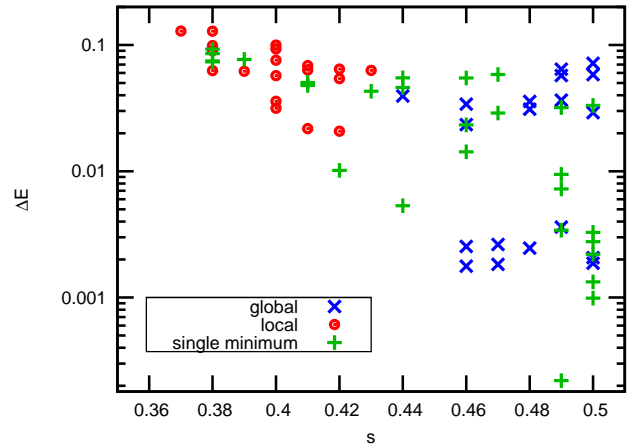}
\caption{(Color online) A scatter plot of the minimum gap for all 47 instances
for size $N=160$ for the 3-regular Max-Cut
problem. For the 19 instances with two minima in the range of
$s$ studied, both are shown, that closer to $s = 0.36$ being denoted
``local'', and the smaller one, at larger $s$, being denoted ``global''. Note
the large scatter in the values of the minimum gap for different instances.}
\label{fig:scatter}
\vspace{-0.7cm}
\end{center}
\end{figure}

So far we have plotted results for the \textit{median} minimum gap, which is
a measure of the \textit{typical} value. However, it is important to note
that there are large
fluctuations in the value of the minimum gap between instances. This is
illustrated in Fig.~\ref{fig:scatter} which presents the values of the minimum
gap for all 47 instances for $N = 160$. For the 19 instances with two minima
in the range of $s$ studied, the minimum at larger $s$ is lower than the one
at smaller $s$.  For these instances, the figure shows both
the ``local'' (smaller-$s$), and the 
``global'' (larger-$s$) minima. 


From Fig.~\ref{fig:scatter} we note that a substantial fraction of instances for
N = 160 have a minimum gap which is
\textit{much} smaller than the median, $0.0291(49)$. Smaller sizes do not have
such a pronounced tail in the distribution for small gaps. 
We should mention that the gap is not precisely determined if 
it is extremely small because we require the condition $\beta \Delta E \gg 1$.
For $N=160$ we took $\beta = 2048$ so this condition is well satisfied for
gaps around the median. Hence we are confident that the median is
accurately determined. However, it is not well satisfied for the smallest gaps
in Fig.~\ref{fig:scatter}. Thus, while it is clear that a non-negligible fraction
of instances for $N=160$ do have a very small minimum gap, the precise value of the
very small gaps in Fig.~\ref{fig:scatter} is uncertain. We note that
if the fraction of instances with a very small minimum gap
continues to increase with $N$, then, asymptotically,
the median would decrease \textit{faster} than that
shown by the fit in the main part of Fig.~\ref{fig:maxcutDE}. 

\section{\label{sec:conclusions}Summary and conclusions}

It was demonstrated in Ref.~\cite{jorg:10} that the quantum adiabatic
algorithm fails to solve random instances of 3-regular 3-XORSAT in polynomial
time, due to an exponentially small gap in the interpolating Hamiltonian which
occurs near $s = s_c = \frac{1}{2}$. This exponentially small gap is associated
with a first order quantum phase transition in the ground state. In this work
we have provided additional numerical evidence for this.  We have also
demonstrated using a duality transformation that the critical value of the
parameter $s$ is in fact at exactly $s_{c}=\frac{1}{2}$. We have also shown that the ground state energy
of the three regular 3-XORSAT model with a transverse field agrees very well
with a replica symmetric (RS) cavity calculation. This provides support for the
claim of Ref.~\cite{jorg:10} that the RS calculation is exact for the
thermodynamic properties of this model.

For the random ensemble of Max-Cut instances that we consider, we find that
the interpolating Hamiltonians exhibit a second order, continuous phase
transition at a critical value $s = s_c \simeq 0.36$. Near this critical value of $s$ we
find that the eigenvalue gaps decrease only polynomially with the number of
bits. However, we also observe very small gaps at values $s>s_{c}$,
i.e.\ in the spin glass phase. An analysis of the fits indicates that a
gap decreasing exponentially with size is preferred over a polynomially
varying gap,
though a stretched exponential fit is also satisfactory.

For both of the problems we studied, the adiabatic interpolating Hamiltonians
are stoquastic. This makes it possible for us to numerically investigate the
performance of the QAA using quantum Monte Carlo simulation and the quantum
cavity method. However it is possible that quantum adiabatic algorithms with
stoquastic interpolating Hamiltonians are strictly less powerful than more
general quantum adiabatic algorithms.

The QMC calculations consider
only instances in which the
problem Hamiltonian, $H_P$, has a doubly degenerate ground state
and a specified value of the ground state energy.
These instances are exponentially rare. By
contrast the cavity approach considers random instances. However, we do not think that these restrictions invalidate the conclusions on the minimum gap summarized in the previous paragraph.

Also, it should be noted that inside the spin-glass phase, QMC techniques
become less and less efficient as the adiabatic parameter $s$ approaches $1$,
i.e.\ when the Hamiltonian approaches the classical problem Hamiltonian. Hence
we have not been able to study $s$ values much larger than 0.5 for a broad range of
sizes. It is possible, indeed likely, that there are
other avoided crossings in this range which might
lead to even smaller minima than those found in the studied range,
$0.3 \leq s \leq 0.5$.  These minima will however not alter the conclusion
that the overall scaling of the running time of the QAA -- when applied to the
Max-Cut problem -- appears to grow exponentially (or perhaps in a stretched
exponential manner) with problem size. 

The first order phase transition in 3-regular 3-XORSAT prevents the quantum
adiabatic algorithm from successfully finding a satisfying assignment. In
contrast, the second order phase transition in 3-regular Max-Cut does not
determine the performance of the quantum adiabatic algorithm on this problem.
In this example the small gaps which occur beyond $s_{c}$ cause the quantum
adiabatic algorithm to fail. These small gaps may be associated with
{}``perturbative crossings'' as described in Refs.~\cite{altshuler:09b,amin:09,altshuler:09,farhi-2009,FSZ10}.

\begin{acknowledgments}
APY and IH acknowledge partial support by the National Security Agency (NSA)
under Army Research Office (ARO) contract number W911NF-09-1-0391, and in part
by the National Science Foundation under Grant No.~DMR-0906366. APY and IH
would also like to thank Hartmut Neven and Vasil Denchev at Google for
generous provision of computer support. They are also grateful for computer
support from the Hierarchical Systems Research Foundation. AS acknowledges support by the National Science Foundation under Grant No. DMR-1104708.  
EF, DG and FZ acknowledge financial support from
MIT-France Seed Fund/MISTI Global Seed Fund grant: ``Numerical simulations and quantum algorithms''.
FZ would like to thank F.~Krzakala, G.~Semerjian and L.~Foini for useful
discussions. DG acknowledges partial support from NSERC.

\end{acknowledgments}

\bibliography{refs,FZrefs,notes} 

\appendix

\section{\label{app:details_QMC}Details of the Quantum Monte Carlo Simulations}

In the Max-Cut problem, the global updates are achieved by dividing the
configurations of the system produced within the SSE scheme into clusters and
then flipping a fraction of them within each sweep of the
simulation~\cite{sandvik:03}.  An important bonus of these cluster updates is
the existence of ``improved estimators'' with which one considers all possible
combinations of flipped and unflipped clusters.  Improved estimators are very
beneficial for determining time-dependent correlation functions as the
signal to noise is much better than with conventional measurements.
It is important to note, however,
that partitioning the SSE configuration into clusters
tends to be very inefficient as the adiabatic parameter $s$ approaches
$1$, where the entire configuration tends to form one big cluster.

A difficulty arises in extracting the gap for the Max-Cut problem due to the
bit-flip symmetry of the Hamiltonian for the following reason.
Eigenstates of the Hamiltonian
are either even or odd under this symmetry (in particular, the ground
state is even). In the $s \to 1$ limit, states occur in even-odd pairs with an
exponentially small gap (see Fig.~2 of Ref.~\cite{hen:11} for an
illustration).
Therefore, the quantity of interest is the gap to the first {\it even} excited
state.
We consider correlation functions of even quantities, so there are only matrix
elements between states of the same parity.  However, the lowest odd level
becomes very close to the ground state near where the gap to the first even
excited state has a minimum. Hence this lowest
odd state becomes thermally populated, with the result that odd-odd gaps are
present in the data as well.

We eliminate these undesired contributions by projecting into the symmetric
subspace of the Hamiltonian.  A way of implementing this projection at zero
temperature is as follows.  In standard quantum Monte Carlo simulations one
imposes {\em periodic} boundary conditions in imaginary time $\tau$ at $\tau =
0$ and $\beta$. To project out the symmetric subspace one imposes, instead,
{\em free} boundary conditions at $\tau = 0$ and $\beta$. The properties of
the symmetric subspace can then be obtained, for $\beta \to \infty$, by
measurements far from the boundaries.  We have incorporated this idea into the
SSE scheme, and use this modified algorithm in the simulations of the Max-Cut
problem. For the convenience of the reader, this idea is explained in greater
detail in Appendix~\ref{app:projSSE}.

For 3-XORSAT there is no need to employ a projection method
because the Hamiltonian does not have bit-flip symmetry.
However, the presence of 3-spin interactions in the problem
Hamiltonian leads to a different difficulty.  In the Max-Cut case, the
(two-spin) Ising interactions allow the SSE configurations to be partitioned
into mutually-exclusive clusters which in turn enable the use of improved
estimators. The nature of the interactions in the 3-XORSAT problem does not
allow for such convenient scheme.  Rather,
construction of clusters for 3-XORSAT is done by
repeatedly attempting to construct clusters using randomly-chosen pairs of
spins taken from the three-spin operators.
The resulting clusters therefore have a random component and can in general
overlap one another.  Moreover, it is important to note that not all attempts
to construct clusters will be successful as in some cases the third spin of
the operators involved may interfere in the construction.  In such cases the
cluster construction must be aborted and restarted.  

\section{\label{app:projSSE}An SSE-based projection-QMC method} 

Here we derive in some detail an SSE-based projection QMC method.
This appendix is rather technical in nature and so is intended mainly 
for readers who are already
familiar with the SSE method.  The purpose
of the algorithm outlined here is to obtain the zero-temperature properties of a system
described by a Hamiltonian $\hat{H}$ projected onto an invariant subspace of a
discrete symmetry operation respected by the Hamiltonian.  In this paper, we
apply the method to $N$ spin-$1/2$ particles and the Max-Cut Hamiltonian,
where the goal is to sample only states which are even under bit-flip
symmetry.  It should be noted, however, that the method is easily generalizable
to other cases.

The success of the projection method follows from the following observation:
Consider the state
\begin{eqnarray}
| \phi \rangle = \left( 2^N\right)^{-1/2} \sum_{\{ z \}} | z \rangle \,, 
\end{eqnarray}
which is a superposition of all $2^N$ basis states $\{ z \}$ with equal amplitude.
Here, the orthonormal set $\{ z \}$ denotes the basis of all classical spin configurations along the $z$-direction. 
Clearly, the state $| \phi \rangle$ is symmetric (even) under bit-flip symmetry.
Acting $\beta$ times with the operator $\e^{-\hat{H}/2}$ 
on the state $| \phi \rangle$, where $\beta$ is a large integer, the resulting state
is, up to a normalization constant, the (even) ground state, i.e.
\begin{eqnarray}
| 0 \rangle \propto | \tilde{0} \rangle \simeq \e^{-\beta \hat{H} /2} | \phi \rangle \,,
\end{eqnarray}
where $| \tilde{0} \rangle$ is the unnormalized ground state.
Note that the bit-flip symmetry shared by both the Hamiltonian $\hat{H}$ and the state $| \phi \rangle$ 
confines the projected states to the even subspace.
This method can be very easily generalized for other systems that respect
other symmetries if one chooses the state $| \phi \rangle$ appropriately.

Next we define a fictitious `partition function' for the scheme,
\begin{eqnarray}
Z  &=& \langle \tilde{0} | \tilde{0} \rangle = \langle \phi |  \e^{-\beta \hat{H} /2} \times  \e^{-\beta \hat{H} /2} | \phi \rangle \\\nonumber
&=& \sum_{z_1,z_2} \sum_{n_1,n_2} \frac1{n_1! n_2!} \left( \frac{\beta}{2}\right)^{n1+n2}
\langle z_1 | (-\hat{H})^{n_1} (-\hat{H})^{n_2} | z_2 \rangle \,.
\end{eqnarray}
As will be immediately clear, the above `partition function' merely serves here as a normalizing factor for the various measured quantities. 

As with the usual SSE approach, we divide the Hamiltonian into components, commonly referred to as `bond' operators,
\begin{eqnarray}
-\hat{H} = \sum_b \hat{H}_b \,.
\end{eqnarray}
The bond operators should obey $\hat{H}_b | z\rangle =h(b,z) |z'\rangle$ for all states in $\{z\}$. Here, $h(b,z)$ is a real number that depends 
in general on the bond index $b$ and the state $z$. The resulting state $|z'\rangle$ must also be one of the $2^N$ basis states chosen for this problem. 
The partition function then becomes
\begin{eqnarray}
Z  = \sum_{z_1,z_2} \sum_{n_1,n_2} \sum_{\{ \hat{S}_{n_1}, \hat{S}_{n_2}\}}\frac1{n_1! n_2!} \left( \frac{\beta}{2}\right)^{n1+n2}
\langle z_1 | \hat{S}_{n_1} \hat{S}_{n_2} | z_2  \rangle \,,
\nonumber \\
\end{eqnarray}
where $\hat{S}_{n_1}$ ($\hat{S}_{n_2}$) denote products or `sequences' of bond operators of length $n_1$ ($n_2)$
and the summation $\{\hat{S}_{n_1}\}$ ($\{\hat{S}_{n_2}\}$) is taken over all possible such sequences.

We next define a new index variable $n=n_1+n_2$, in terms of which the partition function may be written as
\begin{eqnarray} \label{eq:z}
Z  = \sum_{z_1,z_2} \sum_{n}  \sum_{\{ \hat{S}_{n}\}} \frac{\beta^n}{n!} \sum_{n_1=0}^n w(n,n_1)
\langle z_1 | \hat{S}_{n}^{(n_1)} | z_2 \rangle \,,
\end{eqnarray}
where $\hat{S}_{n}^{(n_1)}$ denotes an operator sequence of length $n$ with an
imaginary `cut' running through it, separating the first $n_1$ operators in the
sequence from the last $(n-n_1)$ operators. Here, we have also defined the
weights 
\begin{equation}
w(n,n_1)=2^{-n}{n\choose n_1}
\label{weights}
\end{equation}
which obey 
\begin{eqnarray}
\sum_{n_1=0}^{n} w(n,n_1)=1\,, \quad \forall n\,.
\end{eqnarray}
Before moving on, let us first denote by $| \alpha(n_1) \rangle $ the
(normalized) state obtained by acting with the first $n_1$ operators in the
sequence $\hat{S}_n$ on state $|z_1\rangle$. In particular $|\alpha(0)\rangle
=|z_1\rangle$ and $|\alpha(n)\rangle= |z_2\rangle \sim \hat{S}_n |
z_1\rangle$, where $n$ is the number of operators in the sequence. 

The above expression for the partition function
Eq.~(\ref{eq:z}) has a form very similar to the one obtained in the usual
SSE decomposition. There are however a couple of notable exceptions: (i) While in
the usual SSE the boundaries in the imaginary time direction are periodic (the
requirement $|z_1 \rangle = | z_2 \rangle$ is enforced), here the boundary
conditions are free and the states $| z_1 \rangle$ and $| z_2 \rangle$ are
different in general and are summed over independently.
(ii) To each level along the
operator sequence there is an assigned weight, reflecting the fact that the
different time slices in the `level' direction are not equally weighed. The
time slices in the middle are weighted more than those close to the boundaries. 

\subsection{The updating mechanism}
As in the usual SSE routine, a configuration is
described by the pair $\{ |z_1\rangle,
\hat{S}_n \}$, i.e., a basis state and an operator sequence.  Importance
sampling of the configurations can be done here in much the same way as in the
usual SSE algorithm.  One can use the same local `diagonal' updating steps
by sweeping serially through the sequence $\hat{S}_n$ replacing identity
operators with diagonal ones with appropriate acceptance probabilities and
vice versa.  The acceptance ratios here are exactly the same as those in
the usual SSE procedure. 

The global non-diagonal updates (normally loop or cluster constructions) will
also be the same albeit with one exception. Here, since the boundaries in the
imaginary-time direction are free rather than periodic, loops or clusters
cannot cross the initial and final levels to the other side but must terminate
at the boundaries.

\subsection{Static measurements}
The expectation value of an operator $\hat{A}$ is given by
\begin{align}
\langle \hat{A} \rangle  &= \frac{\langle \tilde{0} | \hat{A} | \tilde{0}
\rangle}{\langle \tilde{0}  | \tilde{0} \rangle} =
{1 \over Z} \langle \phi | e^{-\beta \hat{H}/2} \hat{A} e^{-\beta \hat{H}/2} | \phi
\rangle \\\nonumber = {1 \over Z} &
\sum_{z_1,z_2} \sum_{n, \{ \hat{S}_{n}\}} \frac{\beta^n}{n!} \sum_{n_1=0}^{n}  w(n,n_1)
\langle z_1 | \hat{S}_{n_1} \hat{A} \hat{S}_{n-n_1}  | z_2 \rangle \,,
\label{average_A}
\end{align}
where $\hat{S}_{n_1} \hat{A} \hat{S}_{n-n_1}$ stands for the operator
$\hat{A}$ sandwiched between two parts of the sequence $\hat{S}_n$ splitting
it in two at the cut $n_1$.  The subscripts denote the sizes of each of the two
sequences, $n_1$ and $n-n_1$, respectively. 

For a diagonal operator, whether a bond operator or not, we get
\begin{align}
\langle \hat{A} \rangle  &= \\\nonumber {1 \over Z}
& \sum_{z_1,z_2} \sum_{n, \{ \hat{S}_{n}\}} \frac{\beta^n}{n!}
\sum_{n_1=0}^{n}  \left[ w(n,n_1) a(\alpha_{n_1}) \right]
\langle z_1 | \hat{S}_{n}^{(n_1)} | z_2 \rangle \,,
\end{align}
where $a(\alpha_{n_1})= \langle \alpha(n_1) | \hat{A} | \alpha(n_1) \rangle$.
This means that the expectation value of
$\hat{A}$ will be determined from
\begin{equation}
\langle \hat{A} \rangle =\langle \sum_{n_1=0}^{n}  \left[ w(n,n_1) a(\alpha_{n_1}) \right] \rangle \,.
\label{diag_op}
\end{equation}

As in the usual SSE scheme we can only calculate expectation values of
off-diagonal operators if they are bond operators (or products of bond
operators). A general expression for the average of a bond operator $\hat{A}$
(either diagonal or off-diagonal) is 
\begin{widetext}
\begin{equation}
\langle \hat{A} \rangle  = Z^{-1} \times
\sum_{z_1,z_2} \sum_{n, \{ \hat{S}_{n}\}} \frac{\beta^n}{n!}
\sum_{n_1=0}^{n}  w(n,n_1)
\times \langle z_1 | \hat{S}_{n_1} \hat{S}_{n-n_1+1}  | z_2 \rangle
\times \delta_{\hat{A},\hat{S}^{(1)}_{n-n_1+1}}\,,
\end{equation}
where $\delta_{\hat{A},\hat{S}^{(1)}_{n-n_1+1}}$ means that if the
first operator in $\hat{S}^{(1)}_{n-n_1+1}$ is anything other than $\hat{A}$
then the corresponding weight is zero (this is completely analogous to the corresponding derivation in the usual SSE scheme, see, e.g., \cite{sandvik:92}).  
Making the substitution $n \to n-1$, we arrive at:
\begin{equation}
\langle \hat{A} \rangle  = Z^{-1} \times \sum_{z_1,z_2} \sum_{n, \{ \hat{S}_{n}\}} \frac{\beta^{n-1}}{(n-1)!} 
\times \sum_{n_1=0}^{n-1}  w(n-1,n_1)
\langle z_1 | \hat{S}_{n_1} \hat{S}_{n-n_1}  | z_2 \rangle \times \delta_{\hat{A},\hat{S}^{(1)}_{n-n_1}}
\,.
\end{equation}
Rewriting the above expression gives
\begin{equation}
\langle \hat{A} \rangle  = Z^{-1} \times \sum_{z_1,z_2} \sum_{n, \{ \hat{S}_{n}\}} \frac{\beta^{n}}{n!} 
\times \sum_{n_1=0}^{n-1}  \left( w(n-1,n_1)  \frac{n}{\beta} \delta_{\hat{A},\hat{S}^{(1)}_{n-n_1}} \right)
\langle z_1 | \hat{S}_{n_1} \hat{S}_{n-n_1}  | z_2 \rangle \,,
\end{equation}
which eventually becomes our final expression for the average of a bond
operator:
\begin{equation}
\langle \hat{A} \rangle = \frac{2}{\beta} \langle \sum_{n_1=0}^{n-1}  \left[ w(n,n_1) \left( (n-n_1)
\delta_{\hat{A},\hat{S}^{(1)}_{n-n_1}}\right) \right] \rangle
=\frac{2}{\beta} \langle \sum_{n_1=0}^{n-1}  \left[ w(n,n_1) \left( (n-n_1)
\delta_{\hat{A},\hat{S}^{(n-n_1+1)}_{n}}\right) \right] \rangle
 \,.
\label{bond_op}
\end{equation}
For diagonal bond operators one can use either Eq.~\eqref{diag_op} or \eqref{bond_op}.
For products of bond operators, we similarly get
\begin{equation}
\langle \prod_{i=1}^{m} \hat{A}_i \rangle = 
\left( \frac{2}{\beta}\right)^m   \times
\langle \sum_{n_1=0}^{n-m}  \left[ w(n,n_1)
\left( \frac{(n-n_1)!}{(n-n_1-m)!} \delta_{\prod_{i=1}^{m}
\hat{A}_i,\hat{S}^{(n_1+1..n_1+m)}_{n}}\right) \right] \rangle \,.
\end{equation}

\subsection{Correlation functions}

For correlation-function measurements, let us consider the following expectation value:
\begin{align}
\langle \hat{A}_1(\tau/2) \hat{A}_2(-\tau/2) \rangle &=
\langle \e^{\hat{H} \tau /2} \hat{A}_1 \e^{-\hat{H} \tau} \hat{A}_2 \e^{\hat{H} \tau /2}
\rangle \approx \langle 0 |\e^{\hat{H} \tau /2} \hat{A}_1 \e^{-\hat{H} \tau}
\hat{A}_2 \e^{\hat{H} \tau /2} |0\rangle \\\nonumber
&= \langle 0 | \hat{A}_1 | 0 \rangle \langle 0 | \hat{A}_2 | 0 \rangle +\sum_{m=1}
\langle 0 | \hat{A}_1 | m \rangle \langle m | \hat{A}_2 | 0 \rangle \e^{-(E_m-E_0) \tau} \,.
\end{align}
In our case
\begin{equation} 
\langle \hat{A}_1(\tau/2) \hat{A}_2(-\tau/2) \rangle =
\langle \e^{\hat{H} \tau /2} \hat{A}_1 \e^{-\hat{H} \tau} \hat{A}_2 \e^{\hat{H} \tau /2} \rangle
\sim  \langle \phi | \e^{-\frac1{2}(\beta-\tau) \hat{H} } \hat{A}_1 \e^{-\hat{H} \tau} \hat{A}_2 \e^{-\frac1{2}(\beta-\tau) \hat{H} } | \phi \rangle \,,
\end{equation}
which becomes
\begin{equation}
\langle \hat{A}_1(\tau/2) \hat{A}_2(-\tau/2) \rangle = Z^{-1} \times \sum_{z_1,z_2} \sum_{n, \{ \hat{S}_{n}\}} \sum_{m ,\{ \hat{S}_{m}\}} \frac{(\beta-\tau)^n}{n!} \frac{\tau^m}{m!} \sum_{n_1=0}^{n}  w(n,n_1)
\langle z_1 | \hat{S}_{n_1} \left( \hat{A}_1\hat{S}_m \hat{A}_2\right)  \hat{S}_{n-n_1}  | z_2 \rangle \,.
\end{equation}
For diagonal operators, this can be expressed as
\begin{equation}
\langle \hat{A}_1(\tau/2) \hat{A}_2(-\tau/2) \rangle = 
\langle \sum_{m=0}^{n}  {n \choose m} (\frac{ \tau}{\beta})^m (1-\frac{ \tau}{\beta})^{n-m} \left( \sum_{n_1=0}^{n-m} 
 \left[ w(n-m,n_1) a_1(n_1) a_2(n_1+m)  \right] \right)  \rangle 
\,,
\end{equation}
with an analogous expression for bond correlation functions:
\begin{align}
&\langle \hat{A}_1(\tau/2) \hat{A}_2(-\tau/2) \rangle  =\\\nonumber
&\langle \sum_{m=0}^{n-2}  {n \choose m} (\frac{ \tau}{\beta})^m (1-\frac{ \tau}{\beta})^{n-m} \times \frac{(n-m)(n-m-1)}{(\tau-\beta)^2}
\left( \sum_{n_1=0}^{n-m-2}  \left[ w(n-m-2,n_1) \delta_{\hat{A}_1,\hat{S}^{(n_1+1)}_{n}} 
\delta_{\hat{A}_2,\hat{S}^{(n_1+m+2)}_{n}}
\right] \right) \,.
\end{align}

\subsection{Integrated susceptibilities}

Integrated susceptibilities are given by
\begin{equation}
\int_{0}^{\beta} \rmd \tau \langle \hat{A}_1(\tau/2) \hat{A}_2(-\tau/2) \rangle
= \langle \frac{\beta}{n+1} \sum_{m=0}^{n} \sum_{n_1=0}^{n-m}   a_1(n_1) a_2(n_1+m)   w(n-m,n_1)  \rangle \,,
\end{equation}
for diagonal operators and by
\begin{equation}
\int_{0}^{\beta} \rmd \tau \langle \hat{A}_1(\tau/2) \hat{A}_2(-\tau/2) \rangle
= \langle \frac{n}{\beta} \sum_{m=0}^{n-2} \sum_{n_1=0}^{n-m-2}  w(n-m-2,n_1) \delta_{\hat{A}_1,\hat{S}^{(n_1+1)}_{n}} 
 \delta_{\hat{A}_2,\hat{S}^{(n_1+m+2)}_{n}} \rangle \,,
\end{equation}
for  bond operators. 
\end{widetext}

\subsection{Binomial distribution of the level weights and the large $\beta$ limit}

Note that since the weights assigned to the levels, $w(n, n_1)$ given in
Eq.~\eqref{weights}, correspond to a
binomial distribution with $p=q=1/2$, the weights are sharply peaked
around $n_1=n/2$, i.e., the mid-point of the sequence. More importantly, the
standard deviation of the distribution is $\sigma=(1/2)\sqrt{n}$ meaning that in
the limit of very large $n$, most of the weight is sharply peaked around $n/2$
and there is no need to perform measurements over the entire sequence,
as most of the weight is concentrated in the region within of order $\sqrt{n}$ of $n/2$.

We should emphasize, however,
that this binomial distribution is only needed to reproduce
the precise
average in Eq.~\eqref{average_A} for a specific value of $\beta$. However, $\beta$
does not correspond to a true inverse temperature and the average in
Eq.~\eqref{average_A} does not, in general,
correspond to a Boltzmann distribution. Only for the special case of $\beta
\to \infty$, in which limit the method projects out the ground state, does 
this technique give a valid thermal average. In the case of large $\beta$ we can,
in fact,
obtain the ground state by sampling \textit{anywhere} far from the boundaries. For
example we can obtain $\langle A \rangle$ by the following generalization of
Eq.~\eqref{average_A},
\begin{equation}
\langle A \rangle = {1 \over Z} \langle \phi
| e^{-(1-\lambda) \beta \hat{H}} \hat{A} e^{-\lambda\beta \hat{H}} | \phi \rangle
\, ,
\end{equation}
where $\lambda$ can take any value between 0 and 1 for which both
$\e^{-\lambda \beta \hat{H}} | \phi \rangle$ and
$\e^{-(1-\lambda) \beta \hat{H}} | \phi \rangle$ project out the ground state.
Repeating the above analysis the weights are now sharply peaked around $n_1 =
\lambda n$. Since different values of $\lambda$ give the same result, we can
average over $\lambda$, and hence obtain ground state
properties by omitting the weights $w(n, n_1)$ and averaging \textit{uniformly} over a
range of levels around the middle (in practice we take the middle
$n/4$ levels). Averaging in this way over a range of levels proportional to $n$
(rather than $\sqrt{n}$) improves the signal to noise.

\section{The quantum cavity method for two-Local transverse field spin Hamiltonians}
In this section we motivate and describe the equations which we have solved numerically in our study of random
3 regular Max-Cut. We first derive the cavity equations for a transverse field
spin Hamiltonian with two local interactions on a finite tree. We then briefly
mention the procedure that is used to investigate the infinite
size limit for homogeneous Hamiltonians defined on random regular graphs
(homogeneity means that the interaction is the same on each edge of the
graph). 

\subsection{The quantum cavity method on a tree}

We now review the quantum cavity equations in the continuous imaginary time
formulation~\cite{krzakala:08}.  We consider transverse field spin
Hamiltonians of the form
\begin{equation}
H(\lambda)=H_{0}-\lambda\sum_{i=1}^{N}\sigma_{x}^{i} \label{cavity_ham}
\end{equation}
where $H_{0}$ is diagonal in the Pauli basis and is two local, that is \[
H_{0}=\sum_{(i,j)\in T}H_{ij}\]  where $H_{ij}$ only acts nontrivially on
spins $i$ and $j$ and the graph of interactions $T$ is a tree. 

The starting point for the quantum cavity method is the path integral
expansion of the partition function $\text{Tr}\left[ e^{-\beta H }\right]$,
where $\beta$ is the inverse temperature. This leads to an expression of the
form
\begin{equation}\label{Z_path}
\text{Tr}\left[ e^{-\beta H }\right]=\sum_{\text{paths} P} \tilde{\rho}(P),
\end{equation}
where $\tilde{\rho}$ is a positive function on paths in continuous imaginary
time. A path $P$ can be specified by a number of flips $r$ a sequence of bit
strings $\{z_{1},z_{2},z_{3},...,z_{r+1}=z_{1}\}$ where $z_{i+1}$ differs from
$z_{i}$ by a single bit flip, and an ordered list of times $\{t_{1},t_{2},...,t_{r}\}$ at which
transitions occur.  By normalizing $\tilde{\rho}$ we get a probability
distribution $\rho$ over paths:
\[ \rho(P)=\frac{1}{Z(\beta)}\lambda^{r}dt_{r}dt_{r-1}...dt_{1}e^{-\int_{0}^{\beta}\langle P(t)|H_{0}|P(t)\rangle dt}.\]

A path $P$ of $N$ spins can also be specified as a collection of $N$ one-spin
paths $P^{(i)}$ for $i\in\{1,...,N\}$ where $P^{(i)}$ is specified by $r(i)$
(the number of transitions in the path of the $i$th spin), a single bit
$b(i)\in\{0,1\}$ which is the value taken by the spin at time $t=0$, and a
list of transition times $\{t_{1}^{(i)},t_{2}^{(i)},...,t_{r(i)}^{(i)}\}.$
Then we can also write
\begin{equation}
\rho(P)=\frac{1}{Z(\beta)}\left[\prod_{i=1}^{N}
\lambda^{r(i)}dt_{1}^{(i)}dt_{2}^{(i)}...dt_{r(i)}^{(i)}\right]e^{-\int_{0}^{\beta}\langle
P(t)|H_{0}|P(t)\rangle dt}.
\label{eq:pathint}
\end{equation}
The quantum cavity equations allow one to
determine $\mu_{i\rightarrow j}(P^{(i)})$ , the marginal distribution of the
path of spin $i$ when the interaction $H_{ij}$ between spins $i$ and $j$ is
removed from $H$. This marginal distribution is defined through
\[
\mu_{i\rightarrow j}(P^{(i)})=\frac{1}{N_{i\rightarrow j}}\sum_{P_{k}:k\neq
i}\rho(P)e^{\int_{0}^{\beta}\langle P(t)|H_{ij}|P(t)\rangle dt}.\]
where $N_{i\rightarrow j}$ is a normalizing factor.  The quantum cavity
equations are the following closed set of equations for the cavity
distributions $\{\mu_{i\rightarrow j}\}$.

\begin{eqnarray} 
\lefteqn{\mu_{i\rightarrow j}(P^{(i)})}& &\nonumber\\ & & =
\frac{1}{z_{i\rightarrow j}}\Bigg(\left(
\lambda^{r(i)}dt_{1}^{(i)}dt_{2}^{(i)}...dt_{r(i)}^{(i)}\right)\\
& &\sum_{\stackrel{\displaystyle{P^{(k)}:}}{k\in\partial i\setminus j}}
\bigg[\prod_{k\in\partial i\setminus j}\mu_{k\rightarrow i}\left(P^{(k)}\right)\\ 
& &e^{-\int_{0}^{\beta}\langle P(t)|H_{ik}|P(t)\rangle dt}\bigg]\Bigg)\label{eq:recursion}
\end{eqnarray} 
where $z_{i\rightarrow j}$ is a normalizing constant. From the cavity
distributions it is straightforward to compute expectation values of local
operators such as the magnetization or the energy.

\subsection{The thermodynamic limit:
replica symmetric and 1-step replica symmetry breaking cavity equations}
\label{sec:cavity_Max}

The \emph{replica symmetric} (RS) scheme is exact under the assumption that
the measure over paths in Eq.~\eqref{Z_path} is characterized by a single pure state,
and local correlations decay very quickly as a function of distance. In this case the loops 
of the random graph are irrelevant. For a model defined on a regular graph, and
without disorder in the Hamiltonian, 
such as the Max-Cut problem in Eq.~(\ref{eq:hamMaxCut}), the local environment of each
site is identical to all others.
Then, in the thermodynamic limit all the cavity distributions are the same
($\mu_{i\rightarrow j}=\mu$ for all directed edges $i\rightarrow j).$ Roughly
speaking this assumes that a random regular graph is modeled by an ``infinite
tree'' which is obtained by assuming translation invariance for the recursion
in Eq.~\eqref{eq:recursion}. For a 3 regular antiferromagnet, this gives
{
\begin{widetext}
\begin{equation}
\label{RS_equation}
\mu(P^{(0)}) =  \frac{1}{Z}\left(\lambda^{r}dt_{1}dt_{2}...dt_{r}\right)
  \sum_{P^{(1)},P^{(2)}} 
\Bigg[\mu\left(P^{(1)}\right)\mu\left(P^{(2)}\right) 
  e^{-\int_{0}^{\beta}\langle P(t)|\left[\sigma_{0}\sigma_{1}+\sigma_{0}\sigma_{2}\right]|P(t)\rangle dt}\Bigg].
\end{equation}
\end{widetext}
}
One can then attempt to solve for a distribution $\mu$ over paths which
satisfies this recursion. 
Note that if there is disorder in the Hamiltonian, e.g. in Eq.~\eqref{eq:classical_spinglass}, then
the RS cavity method is more complicated and requires the introduction of a distribution of cavity 
distributions~\cite{cavity,laumann:08}. This is not required in our case.

The 1RSB ansatz is the next level of refinement within the cavity
method--it is described in Refs.~\cite{2010arXiv1008.4844Z} and
\cite{MM09} at the classical level and in the quantum case
in Ref.~\cite{2011PhRvB..83i4513F}.  The 1RSB ansatz is exact under the assumption that
in the thermodynamic limit the distribution $\rho$ in Eq.~\eqref{Z_path} for a random regular graph
is a weighted convex combination of distributions $\kappa$ which have very
little overlap (their support is on non-overlapping sets of paths) and are uncorrelated. 
In the 1RSB
cavity method the Parisi parameter $m\in[0,1]$ is used to assign the
``states'' $\kappa$ different weights in the distribution.  By choosing
$m\in[0,1]$ appropriately one obtains the correct weighting corresponding to
the distribution $\rho$. 

In our study of 3-regular Max-Cut we use the 1RSB quantum cavity method with
$m$ fixed to be $0$. This corresponds to the assumption
that each of the distributions $\kappa$ is weighted evenly in the distribution
$\rho$. We made this choice here because it greatly simplifies the computation~\cite{2011PhRvB..83i4513F}
and does not affect much the result for the ground state energy of this particular model~\cite{mezard:03b}.

We therefore do not present the 1RSB in full generality--we now discuss the
1RSB case with $m=0$. To use this method, we solve for a distribution $Q(\mu)$
over marginal distributions $\mu$ which has the property: If $\mu_{1}$ and
$\mu_{2}$ are drawn independently from $Q$, then $\tilde{\mu}$ defined by 
{
\begin{widetext}
\begin{equation}
\tilde{\mu}(P^{(0)})
=\frac{1}{Z}\left(\lambda^{r}dt_{1}dt_{2}...dt_{r}\right)
\sum_{P^{(1)},P^{(2)}}\Bigg[\mu_{1}\left(P^{(1)}\right)\mu_{2}\left(P^{(2)}\right)\\
 e^{-\int_{0}^{\beta}\langle P(t)|\left[\sigma_{0}\sigma_{1}+\sigma_{0}\sigma_{2}\right]
|P(t)\rangle dt}\Bigg].
\label{eq:tildemu}
\end{equation}
\end{widetext}
}
is also distributed according to $Q$.

Since we cannot represent an arbitrary distribution $Q(\mu)$ in a finite
amount of computer memory, we represent the distribution $Q$ by a number
$N_{D}$ of representatives: that is, marginal distributions
$\mu_{1},\mu_{2},...,\mu_{N_{D}}$ which are each assigned an equal weight in
the distribution.  Furthermore, each cavity distribution $\mu$ is stored as a
list of $N_{R}$ representative paths $P^{(1)},P^{(2)},...,P^{(N_{R})}$ which
are given weights in the distribution $w^{(1)},w^{(2)},...,w^{(N_{R})}$ (with
$\sum_{i}w^{(i)}=1).$ 
\subsection{Details of the Quantum Cavity Numerics for Max-Cut}
Our simulation was run on a Sicortex computer cluster in an embarrassingly
parallel fashion. We ran two independent
simulations at each value of $\lambda.$ We have checked our results with a second
independent implementation of the continuous time cavity method. We used
population sizes $N_{D}=200$ and
$N_{R}=15000.$ We found numerically that there is a systematic error
associated with taking $N_{R}$ to be too small and that this error increases
as $\beta$ is increased. We believe that $N_{R}=15000$ is large enough to make
this error small for our simulation at $\beta=4$ (see \cite{gosset11} for more
details).

For unknown reasons our computer code sometimes (primarily at higher
values of the transverse field $\lambda$ and larger values of $N_{R}$)
did not output the data file. This computer bug
did not seem to compromise the results when the output was produced (we checked this by comparing with results from the independent implementation). 
We have only reported data for values of $\lambda$ where \emph{both} independent
simulations at $N_{R}=15000$ outputted data files.

\end{document}